\documentclass[a4paper,fleqn,usenatbib]{mnras}

\usepackage{newtxtext,newtxmath}
\usepackage[T1]{fontenc}
\usepackage{ae,aecompl}
\usepackage{placeins}

\usepackage{graphicx}	

\title[ALMA observations of CS in NGC 1068]{ALMA observations of CS in NGC 1068: chemistry and excitation}

\author[M.~Scourfield et al.]{M.~Scourfield$^{1}$, \thanks{E-mail: matt.scourfield.18@ucl.ac.uk}
S.~Viti$^{1}$,
S.~Garc\'{i}a-Burillo$^{2}$,
A.~Saintonge$^{1}$,
F.~Combes$^{3}$,
A.~Fuente$^{2}$, \newauthor
C.~Henkel$^{4,5}$,
A.~Alonso-Herrero$^{6}$,
N.~Harada$^{7}$,
S.~Takano$^{8}$,
T.~Nakajima$^{9}$,
S.~Mart\'in$^{10,11}$,\newauthor
M.~Krips$^{12}$,
P.~P.~van~der~Werf$^{13}$,
S.~Aalto$^{14}$,
A.~Usero$^{2}$,
K.~Kohno$^{15,16}$
\\
$^{1}$Department of Physics and Astronomy, University College London, Gower St., London, WC1E 6BT, UK\\
$^{2}$Observatorio  Astron\'{o}mico  Nacional  (OAN-IGN)-Observatorio  deMadrid, Alfonso XII, 3, 28014 Madrid, Spain\\
$^{3}$Observatoire de Paris, LERMA, CNRS, 61 Av. de l'Observatoire, 75014 Paris, France\\
$^{4}$Max-Planck-Institut   f\"ur   Radioastronomie,   Auf   dem   H\"ugel   69,53121 Bonn, Germany\\
$^{5}$Astronomy Department, Faculty of Science, King Abdulaziz University, PO Box 80203, Jeddah 21589, Saudi Arabia\\
$^{6}$Centro de Astrobiolog\'ia (CSIC-INTA), ESAC Campus, 28692 Vil-lanueva de la Ca\~nada, Madrid, Spain\\
$^{7}$Academia Sinica Institute of Astronomy and Astrophysics, P.O. Box 23-141, Taipei 10617, Taiwan, ROC\\
$^{8}$Department of Physics, General Studies, College of Engineering, Nihon University, Tamura-machi, Koriyama, Fukushima 963-8642, Japan\\
$^{9}$Institute for Space-Earth Environmental Research, Nagoya University, Furo-cho, Chikusa-ku, Nagoya, Aichi 464-8601, Japan \\
$^{10}$European Southern Observatory, Alonso de C\'ordova, 3107, Vitacura, Santiago 763-0355, Chile \\
$^{11}$Joint ALMA Observatory, Alonso de C\'ordova, 3107, Vitacura, Santiago 763-0355, Chile \\
$^{12}$Institut  de  Radio  Astronomie  Millimétrique  (IRAM),  300  Rue  dela  Piscine,  Domaine  Universitaire  de  Grenoble,  38406  St.  Martind'Hères, France\\
$^{13}$Leiden Observatory, Leiden University, PO Box 9513, 2300 Leiden,The Netherlands\\
$^{14}$Department of Earth and Space Sciences, Chalmers University of Technology, Onsala Observatory, 439 94 Onsala, Sweden\\
$^{15}$Institute of Astronomy, School of Science, The University of Tokyo, 2-21-1 Osawa, Mitaka, Tokyo 181-0015, Japan\\
$^{16}$Research Center for the Early Universe, School of Science, The University of Tokyo, 7-3-1 Hongo, Bunkyo-ku, Tokyo 113-0033, Japan
}

\date{Accepted XXX. Received YYY; in original form ZZZ}

\pubyear{2020}

\begin{document}
\label{firstpage}
\pagerange{\pageref{firstpage}--\pageref{lastpage}}
\maketitle

\begin{abstract}
We present results from Atacama Large Millimeter/submillimeter Array (ALMA) observations of CS from the nearby galaxy NGC 1068 ($\sim14$ Mpc). This Seyfert 2 barred galaxy possesses a circumnuclear disc (CND, $r\sim200$ pc) and a starburst ring (SB ring, $r\sim1.3$ kpc). These high-resolution maps ($\sim0.5''$, $\sim35$ pc) allow us to analyse specific sub-regions in the galaxy and investigate differences in line intensity ratios and physical conditions, particularly those between the CND and SB ring.
Local thermodynamic equilibrium (LTE) analysis of the gas is used to calculate CS densities in each sub-region, followed by non-LTE analysis conducted using the radiative transfer code RADEX to fit observations and constrain gas temperature, CS column density and hydrogen density. Finally, the chemical code UCLCHEM is used to reconstruct the gas, allowing an insight into its origin and chemical history.
The density of hydrogen in the CND is found to be $\geq10^5$ cm$^{-2}$, although exact values vary, reaching $10^6$ cm$^{-2}$ at the AGN. The conditions in the two arms of the SB ring appear similar to one another, though the density found ($\sim10^4$ cm$^{-2}$) is lower than in the CND. The temperature in the CND increases from east to west, and is also overall greater than found in the SB ring.
These modelling methods indicate the requirement for multi-phase gas components in order to fit the observed emission over the galaxy. A larger number of high resolution transitions across the SLED may allow for further constraining of the conditions, particularly in the SB ring.
\end{abstract}



\section{Introduction}
\label{sec:intro}

Galaxy evolution is driven by star formation, and so feedback processes which control the amount of gas available to form stars play an important role \citep[e.g.][]{Saintonge}. This feedback can originate from the active galactic nucleus (AGN) or stellar components of the galaxy, the primary contribution to the latter being massive stars which form in regions of hot, dense gas which are themselves shielded from stellar winds \citep{massive}.

Sulphur-bearing species in particular are used to trace the evolution of hot cores, due to their sensitivity to physical and chemical variations \citep[e.g.][]{massiveModels, Awad, Vidal, Martin05}, and have been observed to be enhanced during massive star formation \citep{Bayet3}. Of these CS serves as one of the best tracers of dense gas due to its high critical density, with the CS(2-1) transition having a $n_{crit}\sim10^5$ cm$^{-3}$ and the CS(7-6) transition $n_{crit}\sim10^7$ cm$^{-3}$ \citep{Plume}\footnote{Models by \citet{Nanase} were able to reproduce CS(2-1) observations in the Orion A \citep{Kauffmann} and Orion B clouds \citep{Pety} at densities lower than the critical density of the transitions, suggesting that CS emission can also come from low-density regions.}.



The rate of massive star formation has been observed to be particularly enhanced in starburst (SB) galaxies, possibly as a result of mergers between galaxies with high amounts of interstellar matter. Of particular interest are SB galaxies in which an AGN is also present. In such galaxies, the interstellar medium (ISM) is affected by a number of complex, high energy processes as a result of the UV and X-ray radiation, cosmic rays and shocks caused by AGN feedback which process the gas and dust in the central regions. Large amounts of dense circumnuclear gas hint at the combined action of both star formation and AGN-driven feedback on the physical properties and chemistry of molecular gas \citep{Viti}.

Emission line observations can be used to trace molecular outflows of galaxies, giving insight into mechanical feedback. Through the use of photon-dominated region (PDR) and X-ray dominated region (XDR) codes to interpret these observations, they can also be used to differentiate between the effects of massive star formation and AGN. Models of AGN/SB systems \citep{Bayet, Bayet2} have not revealed any unique tracer capable of distinguishing these galaxies from those in which only the SB is present.

NGC 1068 is a nearby Seyfert 2 type barred galaxy, which serves as a prominent example of such an AGN/SB galaxy. Due to both its proximity ($\sim14$ Mpc; \citealt{Bland}) and brightness (L$_{\text{Bol}} \sim2.5 - 3.0\times10^{11}$ L$_\odot$; \citealt{Bock}), it has been extensively studied using molecular line observations in efforts to understand the fuelling and subsequent feedback within the central regions of the galaxy. As a result, the galaxy has become a prototype of both Seyfert and AGN/SB galaxies.

Observations of the first two transitions of the CO molecule made using the Plateau de Bure Interferometer (PdBI) by \citet{Schinnerer} revealed the structure of the galaxy to consist of an SB ring of radius $\sim 1-1.5$ kpc with a circumnuclear disk (CND) surrounding the AGN (r $\sim 200$ pc). Subsequent observations of higher CO transitions give agreeing views of the gas distribution \citep[e.g.][]{Tsai, Garcia, Viti, Burillo}.

Molecular line surveys of the galaxy conducted using the NRO 45-m single-dish telescope and Atacama Large Millimeter/submillimeter Array (ALMA) early cycle data have classified the distribution of molecules, and hint at the chemistry present in the CND \citep{Nakajima11, Nakajima15, Nakajima18, Takano14}.

Previous work by \citet{Garcia}, using observations from the ALMA and the PdBI, found a possible footprint of AGN feedback in the form of dramatic changes in line ratios across the CND, tightly correlated with UV and X-ray illumination of the region by the AGN. By analysing the kinematics of the CO gas they also identified an AGN driven outflow, with an outflow rate of $63^{+21}_{-37}~M_{\odot}~yr^{-1}$ from the CND, giving a gas depletion timescale $\leq1$ Myr. Efficient gas inflow from the SB ring however refreshes the reservoir over a much longer timescale.

In the follow up paper \citep{Viti}, chemical analyses of CO, HCO$^+$, HCN and CS molecular observations were conducted within a number of sub-regions of interest in the galaxy. This analysis suggests a pronounced chemical differentiation across the CND as well as a three phase ISM, one of which is a shocked gas phase which gives rise to the high CS abundances and the remainder possessing high ionisation rates, and showed the SB to have a lower molecular content than the CND. Chemical modelling suggests the CND to be cooler in the east, and gives average gas densities across the CND in excess of $10^4$ cm$^{-3}$. The modelling also suggests the chemical composition of the SB ring to be similar to that found in the west of the CND, albeit with lower gas density and temperature.

Other multi transition CS studies in nearby galaxies include \citet{Mauersberger, Bayet4, Bayet3, Martin06, Aladro} which look at gas in the centre of galaxies. \citet{Kelly} use multiple CS observations taken at various offsets within SB galaxies in order to map the conditions within.

In this paper we focus on the chemical analysis of CS observations within NGC 1068, using four CS line transitions with a common spatial resolution of 35 pc. By modelling the emission in various sub-regions and comparing the physical conditions across the outer and inner regions of the galaxy, we aim to disentangle the effects of the AGN and SB on the gas near the centre of the galaxy.

In this work, we present our observations in Sect. \ref{sec:obs}, and the maps produced from these are discussed in Sect. \ref{sec:data}. LTE analysis of the transitions are carried out using both the Boltzmann relation and rotational diagram method in Sect. \ref{sec:LTE}. Modelling of the gas is carried out in Sect. \ref{sec:nLTE} using the radiative transfer and chemical codes RADEX and UCLCHEM, to provide an analysis for the non-LTE case. A summary of this works conclusions is given in Sect. \ref{sec:conc}.

\section{Observations}
\label{sec:obs}

\subsection{ALMA data}

The CS transitions of NGC 1068 used in this paper were observed using ALMA. The CS(7-6) emission was obtained during cycle 0 using band 7 receivers (project-ID: \#2011.0.00083.S; PI: S. Garc\'{i}a-Burillo), and was included in the anlysis by \citet{Garcia} and \citet{Viti}. Additional transitions used in this paper are the cycle 2, band 3 receiver emission of CS(2-1) (project-ID: \#2013.1.00055.S; PI: S. Garc\'{i}a-Burillo), and the CS(3-2) and CS(6-5) emissions obtained in cycle 3, in bands 4 and 7 respectively (project-ID: \#2015.1.01144.S; PI: S. Viti). All of these observations have been primary beam corrected.

The phase tracking centre was set to $\alpha_{2000} =$ \mbox{$02^{\rm h}42^{\rm m}40\fs771$}, $\delta_{2000} =$ \mbox{$-00\degr00\arcmin47\farcs94$} (J2000 reference system, as used throughout the paper) in each case, the position of the galaxy's centre in the SIMBAD Astronomical Database, from the Two Micron All Sky Survey-2MASS survey \citep{2MASS}. This is offset relative to the galaxy AGN at $\alpha_{2000} =$ \mbox{$02^{\rm h}42^{\rm m}40\fs710$}, $\delta_{2000} =$ \mbox{-00$^\circ$:00$'$:47{\hbox{$\,.\!\!^{\prime\prime}$}}94 by $<1''$}, and corresponds to a peak in CO emission \citep{Garcia}. For each spectral setup, a total of four spectral bands were measured.

Initial reduction of the data was carried out using the ALMA reduction package CASA\footnote{version 5.4.0; \url{http://casa.nrao.edu}}  \citep{CASA}, and then exported to GILDAS\footnote{version oct18b; \url{http://www.iram.fr/IRAMFR/GILDAS}} and Python (making use of the Astropy package\footnote{version 3.2.1; \url{https://www.astropy.org}}, \citealt{astropyI, astropyII}) for further reduction. The $1\sigma$ threshold for each map was also obtained in GILDAS by calculating the rms of the signal over an area towards the centre of the galaxy in a channel with no line detection\footnote{These velocity channels were centred on $497$ km/s for CS(2-1), $247$ km/s for CS(3-2), $805$ km/s for CS(6-5) and $-450$ km/s for CS(7-6). The velocity ranges of each channel are given by the resolution of the observation. In the case of CS(2-1), CS(6-5) and CS(7-6) these are sufficiently far enough from the observed position of the line to ensure no emission is present. Due to the limited wavelength range of the CS(3-2) observation this could not be done, and so extra care was taken to ensure the area over which the rms was calculated contained no emission in the channel.}.

As stated in section \ref{sec:intro} we take the distance to NGC 1068 to be $\sim$14 Mpc, such that 1 arcsecond corresponds to 70 pc, and take the systemic velocity of the galaxy as $v_{sys}\text{(HEL)} = 1127$ km s$^{-1}$ \citep{Garcia}. Relative velocities throughout the paper are compared to this value.

Velocity-integrated maps obtained for each of the transitions are shown in Appendix \ref{sec:emission}, with emission clipped below the $3\sigma$ level and outside the field of view (FOV) which is taken as the size of the ALMA primary beam.

\begin{table}
\caption{Names and coordinates of the sub-regions of interest across the CND and SB Ring.}
\label{table:Regions}
\centering
\begin{tabular}{c c c}
\hline\hline
Sub-Region & RA$_{J2000}$ & Dec$_{J2000}$ \\
\hline
E Knot & $02^{\rm h}42^{\rm m}40\fs771$ & $-00\degr00\arcmin47\farcs84$ \\
W Knot & $02^{\rm h}42^{\rm m}40\fs630$ & $-00\degr00\arcmin47\farcs84$ \\
AGN    & $02^{\rm h}42^{\rm m}40\fs710$ & $-00\degr00\arcmin47\farcs94$ \\
CND-N  & $02^{\rm h}42^{\rm m}40\fs710$ & $-00\degr00\arcmin47\farcs09$ \\
CND-S  & $02^{\rm h}42^{\rm m}40\fs710$ & $-00\degr00\arcmin49\farcs87$ \\
NSB-a  & $02^{\rm h}42^{\rm m}40\fs840$ & $-00\degr00\arcmin33\farcs00$ \\
NSB-b  & $02^{\rm h}42^{\rm m}41\fs290$ & $-00\degr00\arcmin34\farcs97$ \\
NSB-c  & $02^{\rm h}42^{\rm m}39\fs820$ & $-00\degr00\arcmin39\farcs26$ \\
SSB-a  & $02^{\rm h}42^{\rm m}40\fs317$ & $-00\degr01\arcmin01\farcs84$ \\
SSB-b  & $02^{\rm h}42^{\rm m}39\fs820$ & $-00\degr00\arcmin52\farcs79$ \\
SSB-c  & $02^{\rm h}42^{\rm m}39\fs870$ & $-00\degr00\arcmin55\farcs32$ \\
\hline
\end{tabular}
\end{table}

\subsubsection{Cycle 0}

During this cycle, CS(7-6) emission (342.883~GHz rest frequency) was observed. A mosaic of eleven fields from band 7 was used to cover both the CND and SB ring, each with a field of view of 17$''$. Four total tracks were observed over the period June to August 2012 with on source times of 24 - 39 minutes each, for a total of 138 minutes. 18 - 27 antennas were used in each track, with projected baselines of 17 - 400 m.

Two of the spectral windows were in the lower sideband (LSB) centred on 341.94 GHz and 343.75 GHz and two in the upper sideband (USB) centred on 353.81 GHz and 355.68 GHz. Each of these windows had a bandwidth of 1.88 GHz and a resolution of 488 kHz ($\sim$ 13 km s$^{-1}$).

The SB ring and the CND of NGC 1068 fall within the central $r=25''$. The field of view of the final mosaic is $\sim50''$, which is sufficiently large to cover these.

The beam size used was $0.59'' \times 0.48'' (\text{PA }61^\circ$), and the absolute error on the flux values was assumed to be 15\%. To convert the map from Jy beam$^{-1}$ to kelvin a conversion factor of 36 K Jy$^{-1}$ beam was used, and the $1\sigma$-rms was found to be 0.04 K.

\subsubsection{Cycle 2}
\label{sect:cycle2}

In order to map CS(2-1) (rest frequency 97.981 GHz) one track was observed in August 2015 with band 3 receivers, using 34 antennas with baselines from 12 - 1430 m and an exposure time of 60 minutes. The resulting 70$''$ FOV was sufficient to cover both the CND and SB ring.

Each spectral window had a bandwidth of 1.87 GHz and a resolution of 977 kHz ($\sim$ 9 km s$^{-1}$). The two LSB windows were centred at 86.3  and 88.1 GHz, and in the USB at 98.2 and 100.1 GHz.

A $0.75'' \times 0.51'' (\text{PA }75^\circ)$ beam was used, with an assumed 10\% absolute flux error. Conversion of the map to K was done with a conversion factor of 340 K Jy$^{-1}$ beam, with a $1\sigma$-rms of 0.14 K.

\subsubsection{Cycle 3}

During June-July 2016, observations were made with both band 4 and 7 receivers using 40 antennas. The band 4 observations of the CS(3-2) transition (rest frequency 146.969 GHz) were taken with an exposure time of 44 minutes and baselines of 15 - 1100 m, using a $0.66'' \times 0.44'' (\text{PA }85^\circ)$ beam centred on 144.6 and 146.4 GHz in the LSB and 156.6 and 158.5 GHz in the USB. Each of these windows has a resolution of 1953 kHz ($\sim$ 4 km s$^{-1}$) and bandwidth of 1.87 GHz.

The resulting map has an FOV of 42$''$ which is comparable in size to the region of the SB ring, such that the SNR begins to drop off immediately outside the SB ring. A conversion factor of 194 K Jy$^{-1}$ beam and an assumed absolute flux error of 15\% were used. The $1\sigma$-rms noise level was 0.05 K.

The band 7 CS(6-5) (rest frequency 293.912 GHz) observations used a $0.50'' \times 0.45'' (\text{PA }81^\circ)$ beam with a 61 minute exposure and baselines of 15 - 772 m to produce a map with a 21$''$ FOV. As this is significantly smaller than the size required to cover the SB ring we exclude the CS(6-5) transition from the analysis of sub-regions within the SB ring. This FOV is however sufficiently large to cover the central r = 3$''$ containing the CND and so detections in this region are still used in the analysis.

The spectral windows in the LSB were centred on 291.87 and 293.74 GHz, and on 303.79 and 305.66 GHz in the USB, with bandwidths of 3.74 GHz and resolutions of 3906 kHz ($\sim$ 4 km s$^{-1}$) in each. The assumed absolute error on flux values was 15\%, and the conversion factor 63 K Jy$^{-1}$ beam. The $1\sigma$-rms was found to be 0.03 K.

Both the band 4 and 7 cycle 3 observations were interpolated to a resolution of $\sim$ 9 km s$^{-1}$ in order to match the cycle 2 observations (see Sect. \ref{sect:cycle2}).

\subsection{Ancillary data}

We also use calibrated images of Pa$\alpha$ line emission in the galaxy, obtained by retrieving the HST NICMOS (NIC3) narrow-band (F187N, F190N) data of NGC 1068 from the Hubble Legacy Archive (HLA). Details on the calibration pipeline used to reprocess these images are given in Sect. 2.2 of \cite{Garcia}. The pixel size of the HLA images is 0{\hbox{$\,.\!\!^{\prime\prime}$}}1 square.

\section{Data maps}
\label{sec:data}

\subsection{Molecular line emission}

\begin{figure*}
	\centering
	\includegraphics[width=17cm]{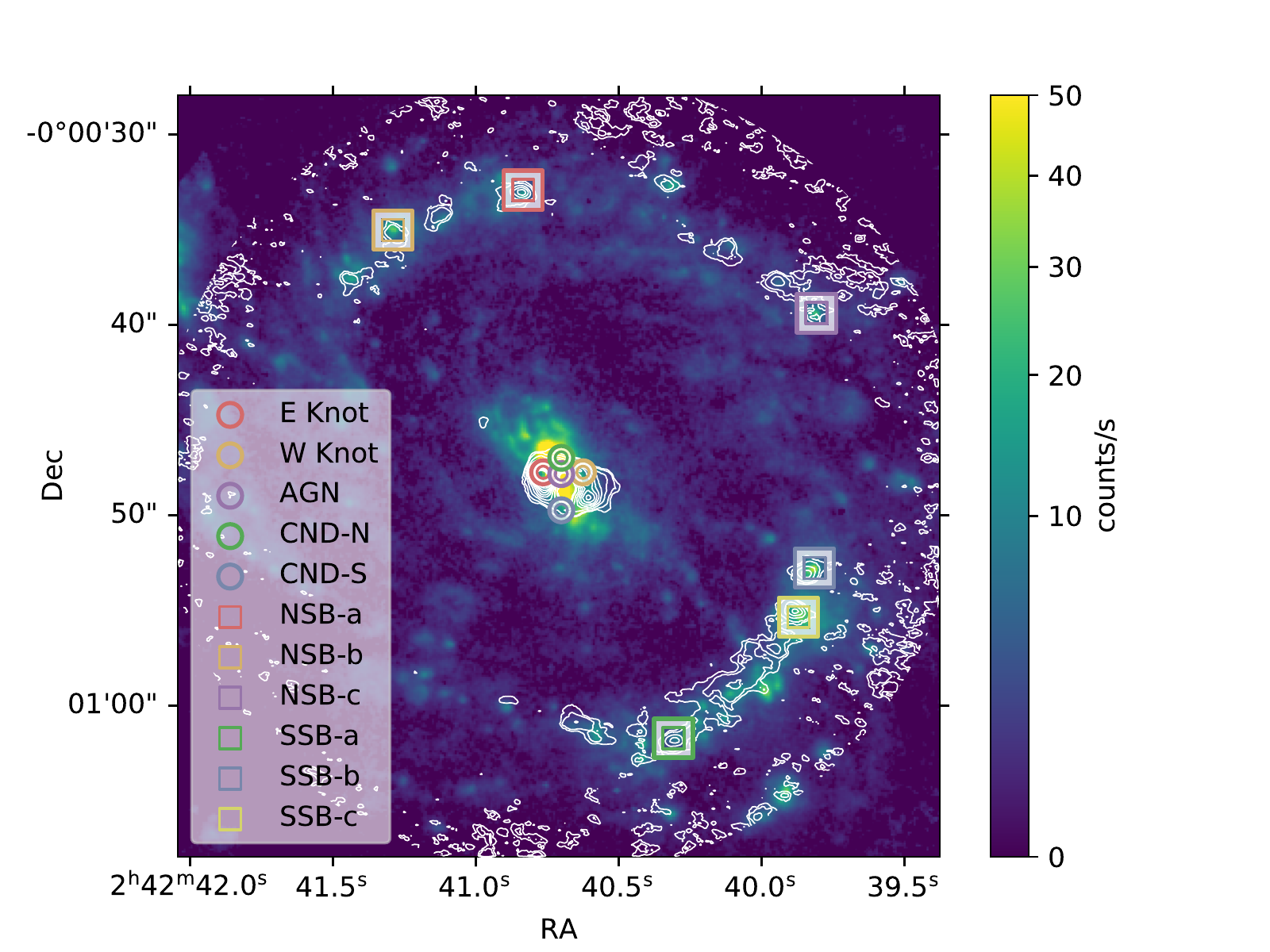}
	\caption{Overlay of the ALMA CS(3-2) emission map (contours) of NGC 1068 onto the HST Pa$\alpha$ map (coloured). The inner, brighter source of Pa$\alpha$ is the galaxy's CND, and the outer ring like structure its SB ring. Between these is a bar, which is largely undetected in these two emitters. The CS contour levels are at $3\sigma$ and $5\sigma$ to $40\sigma$ in intervals of $5\sigma$, where sigma is 9 K km s$^{-1}$.}
	\label{fig:alpha}
\end{figure*}

Figure \ref{fig:alpha} shows the colourmap of Pa$\alpha$ with the CS(3-2) contours overlayed as an example.

The emission in the Pa$\alpha$ map comes from two main regions, an inner and an outer region. These regions are the CND and the SB ring respectively, and are compared throughout this analysis. Between these two regions is a bar connecting the CND and SB ring, which can be seen in CO maps of the galaxy \citep{Garcia}. The absence of this bar region in the CS maps suggests the gas within is less dense than that in the CND and SB ring. Due to this lack of CS detection in the bar region, we do not look at it in this analysis.

Within the CND and SB ring a number of sub-regions of interest were selected, in which detailed analysis of the observations were carried out. The CND sub-regions chosen all correspond to the sub-regions examined in \citet{Viti}, whereas in the SB ring a number of new sub-regions are chosen in addition to the SB ring sub-region from the paper (which has been re-labeled to `SSB-a' in this paper to match the naming convention used for other sub-regions in the ring). These additional sub-regions were chosen (see below) to cover both the north and south arms of the SB ring in order to investigate the conditions over the SB as a whole. The names and coordinates of each of the sub-regions are given in Table \ref{table:Regions}, and the spectra between $\pm 250$ km/s of each CS transition for these sub-regions are shown in Appendix \ref{sec:emission} with Gaussians fitted to the emission in the cases where there is a detection.

The SB ring of NGC 1068 is made up of two spiral arms, both of which subtend an angle of $180^\circ$ over a range $\sim1-1.7$ kpc from the galactic centre. Three sub-regions of interest were identified in each arm. Of these the SSB-a and NSB-a sub-regions correspond to peaks in CO emission \citep{Garcia} and the remaining sub-regions to peaks in Pa$\alpha$. The sub-regions also correspond spatially to peaks in the 349 GHz dust continuum \citep[see][Fig. 1a]{Garcia}.

The SB ring is most clearly seen in the CS(3-2) map, with the emission in the north arm appearing clumpier than that in the south. This is true also for the CS(2-1) map, in which the overall clumpiness of the detections is also greater. This is somewhat surprising, as the CS(2-1) transition should trace less dense gas compared to the 3-2 line. Therefore it is likely that this is a result of the higher noise of the CS(2-1) data ($\sigma_{2-1}$ = 0.14 K compared to $\sigma_{3-2}$ = 0.05 K). In the CS(7-6) map the detection in both arms is reduced to only a small number of clumps, located mainly at the sub-regions selected.

The area of brightest CS emission resides in the CND, an annulus of size $\sim5''\times4'' (350\times280~\text{pc})$ with the major axis at a PA of $\sim45^\circ$ and shifted such that its centre is offset to the south west of the AGN. In the CND the sub-regions of interest include the AGN and a pair of knots within the disk identified from peaks in the CO map, one $\sim 0.9''$ to the east (E Knot) and the other $\sim 1.2''$ to the west (W Knot) of the AGN. The two additional sub-regions, one $\sim 0.8''$ to the north (CND-N) and the other $\sim 1.9''$ to the south (CND-S) of the AGN position, are located near to the contact points of the jet-ISM working interface \citep{Viti}.


In all maps, except in CS(3-2), there is a break in the CND to the south. This break is also seen in CO(6-5), but is absent in CO(3-2) and CO(2-1) \citep{Garcia, Burillo}. An additional break to the north is also present in the CS(2-1) map.

The locations of peaks in the CND are common across all transitions. The strongest peak occurs at the E Knot, and the second strongest at the W Knot. In maps without breaks, these two sub-regions are connected to the north by a bridge of lower emission through the CND-N, and similarly to the south through the CND-S.

\subsection{Molecular line ratios}
\label{sec:rat}

The ratio between lines in a galaxy has been observed to vary with type, especially between AGN and SB dominated galaxies \citep{Sternberg, Kohno01, Kohno03, Kohno05, Krips, Aladro13, Privon15}. As such we also produce line ratio maps of the galaxy in order to see the effect of the AGN on the surrounding gas by comparing the ratios in the CND and SB ring.

Due to the different resolutions of each of the transition maps, it was necessary to degrade them all to the lowest common spatial resolution through convolution with an appropriate Gaussian for each transition.\footnote{These maps were only used to produce ratio maps, the analysis makes use of the original maps.} In this case a resolution of $0.75'' \times 0.51'' (75^\circ)$ was chosen, corresponding to the CS(2-1) map. Three independent CS line ratio maps were thus produced by dividing the flux in corresponding pixels of the two un-clipped emission maps: \mbox{CS(3-2)/CS(2-1)}, \mbox{CS(6-5)/CS(2-1)} and \mbox{CS(7-6)/CS(2-1)}, hereafter referred to as R32/21, R65/21 and R76/21 respectively. These maps are shown in Appendix B.


The small FOV of the CS(6-5) observation means that the R65/21 ratio map does not extend to the SB ring. Of the remaining two maps the structure of the arms is more clearly visible in R32/21, whereas there is significantly less detection in the ring for R76/21.

The approximate locations of peaks and troughs in the CND are common across all of the ratio maps; there are peaks to the south east and south west, and a third to the north west, above the W Knot. The value of each of the ratios drops off at the inner edges surrounding the AGN and also at the outer edges of the CND. These two areas of low values meet between the AGN and the CND-N, forming a strip which divides the east and west of the disk. Another such strip is present to the south of the W Knot.

These drops in ratio signify that the lower J CS emission is dominant over the higher J in these central areas surrounding the AGN.



\section{LTE analysis}
\label{sec:LTE}

Using the fits to the emission lines (shown in appendix \ref{sec:emission}), we are able to calculate the column density of each transition, $N_u$, by using

\begin{equation}
	N_u = \frac{8\pi k\nu^2W}{hc^3A_{ul}}\left(\frac{\Delta\Omega_a}{\Delta\Omega_s}\right)C_\tau, \label{Nu}
\end{equation}

\noindent where $W$ is the integrated line intensity, $A_{ul}$ the Einstein A-coefficient, $C_\tau$ the optical depth correction factor in the case of saturated lines. The bracketed terms $\Delta\Omega_a$ and $\Delta\Omega_s$ are the solid angles of the antenna and the source respectively, and together give the beam filling factor.

In this analysis we take $W$ to be equal to the area of the fitted Gaussian, $1.06\times T_{peak}\Delta\nu$, where $T_{peak}$ is the peak temperature, $\Delta\nu$ the full width at half-maximum (FWHM) in units of velocity and the factor $1.06$ accounts for the geometry. The derived integrated ubintensities and their errors are shown in Table \ref{table:intensities} We also initially assume that the source completely fills the beam such that the beam filling factor is unity and that all emission within a sub-region traces the same gas.

\begin{table*}
\caption{Integrated line intensities (K km/s) derived from fitting parameters for the transitions in each sub-region and their associated errors. Dashes indicate no fit was found for that particular transition in the sub-region. In the case of the SB ring regions, this is because the FOV of the CS(6-5) observation did not cover the ring. Note that the value obtained for CS(7-6) in the NSB-c sub-region is an upper limit.}
\label{table:intensities}
\centering
\begin{tabular}{c c c c c}
\hline\hline
Sub-Region & CS(2-1) & CS(3-2) & CS(6-5) & CS(7-6)\\
\hline
EKnot & $387\pm13$ & $407\pm12$ & $159\pm3$ & $157\pm5$ \\
WKnot & $87\pm11$ & $104\pm6$ & $33\pm2$ & $31\pm4$ \\
AGN & - & $44\pm6$ & $33\pm3$ & $19\pm4$ \\
CNDN & $36\pm11$ & $70\pm4$ & $15\pm2$ & $12\pm4$ \\
CNDS & $25\pm15$ & $34\pm5$ & $10\pm3$ & - \\
NSB-a & $77\pm7$ & $67\pm4$ & - & $30\pm3$ \\
NSB-b & $29\pm6$ & $27\pm3$ & - & $5\pm3$ \\
NSB-c & $14\pm6$ & $17\pm5$ & - & $<2$ \\
SSB-a & $57\pm7$ & $51\pm4$ & - & $7\pm3$ \\
SSB-b & $46\pm6$ & $33\pm3$ & - & $6\pm3$ \\
SSB-c & $41\pm8$ & $27\pm3$ & - & $4\pm2$ \\
\hline
\end{tabular}
\end{table*}

For a region in LTE the rotational temperature, $T$, is equal for all transitions. This allows us to use the densities of individual transitions to calculate the total column density, $N,$ of a species by the relation

\begin{equation}
	N = \frac{N_uZ(T)}{g_ue^{\frac{-E_u}{kT}}} \label{N},
\end{equation}

\noindent where $Z(T)$ is the partition function, which varies with temperature,  and $g_u$ the statistical weight of level $u$, given by $2J + 1$.

Tables \ref{table:N_CND} and \ref{table:N_SB} show the values for $N$ computed from each transition for each of the sub-regions in the CND and SB ring respectively using values from 15 to 250 K for the temperature, following the analysis of \citet{Viti}, in order to cover a variety of conditions. The median value across the entire region is also given in each table.

We have assumed here that the gas is optically thin, setting the optical depth correction factor in Equation \ref{Nu} to unity. In the event this is not true, these values correspond to lower limits for the column densities of the sub-regions.

For a given temperature all of the transitions should theoretically give the same value for the column density in a sub-region; however, this is not the case here. This suggests that at least one of the assumptions made is incorrect.

For temperatures above 15 K, the calculated CS column density for a given region decrease as we look at higher J transitions. The rate of this decrease becomes greater as the temperature increases, up to 150 K, beyond which it does not grow further. The overall drop in the SB ring at high temperatures approaches 2 orders of magnitude for all regions except NSB-a. The overall drop in NSB-a is instead more in line with what is seen in the CND, approaching 1.5 orders of magnitude \footnote{We have assumed that the overall drop in the CND-S column density is similar to that observed in other sub-regions in the CND as the transitions that are detected match the trend seen in these sub-regions; however, due to the non-detection of CS(7-6) it is possible that the actual drop differs.}.

Possible explanations for this decrease include either the density or the kinetic temperature of the gas not being high enough to excite the higher J transitions. The lack of this drop in the 15 K calculations suggests that this is the true kinetic temperature of the gas, supporting the latter reason. This is lower than the temperatures found by \citet[][50 K]{Viti}, however as they only used a single, high-J CS transition it is possible that the it was tracing a higher temperature gas component. Other analyses looking at larger numbers of CS transitions find lower rotational temperatures in line with our conclusion \citep[e.g.][]{Bayet2}.

Among the sub-regions in the CND, the E Knot possesses the highest column densities (up to $10^{16}$) and the CND-S the lowest (with a maximum of $10^{15}$) across all temperatures, though the difference in column densities between the two sub-regions varies by only slightly more than an order of magnitude in all conditions. Assuming CS to be a dense gas tracer, and that the CS abundance in the CND is high, this identifies them as the sub-regions in the CND with the most and least dense gas respectively.

Within the south arm of the SB ring, each of the sub-regions have similar values for the column density at each temperature, within a factor of 2. NSB-a is significantly denser than the other sub-regions in the north arm, over an order of magnitude greater than NSB-c, while the other two sub-regions differ by only a factor $\sim2$.

With the exception of NSB-a, the difference in column densities between the two arms are all with a factor $\sim4$ of each other, suggesting the gas between these sub-regions is largely similar.

Comparing the median CS column densities from the CND and SB ring reveals that values in the CND are greater in all cases. This difference remains approximately the same for a given transition as temperature increases. However, as the rotational quantum number increases so too does the difference, from a factor $\sim1.4$ for CS(2-1) to $\sim4.8$ for CS(7-6), showing that the decrease is greater in the SB ring.

\begin{table}
\caption{Total CS column densities ($\times 10^{14}$ cm$^{-2}$) in NGC 1068 for each of the sub-regions of interest in the CND calculated from individual level densities for various temperatures, assuming LTE, optically thin emission and that the source fills the beam. }
\label{table:N_CND}
\centering
\begin{tabular}{c c c c c c}
\hline\hline
Sub-Region & T (K) & CS(2-1) & CS(3-2) & CS(6-5) & CS(7-6)\\
\hline                                   
EKnot      &         15 & 11     & 8.5     & 8.8     & 19     \\
           &         50 & 31     & 17     & 3.4     & 3.4     \\
           &        100 & 59     & 29     & 4.1     & 3.5     \\
           &        150 & 110     & 56     & 7.0     & 5.6     \\
           &        200 & 170     & 82     & 9.6     & 7.5     \\
           &        250 & 160     & 81     & 9.2     & 7.1     \\
\hline
WKnot      &         15 & 2.5     & 2.1     & 1.8     & 3.7     \\
           &         50 & 7.1     & 4.3     & 0.7     & 0.7     \\
           &        100 & 13     & 7.5     & 0.9     & 0.7     \\
           &        150 & 25     & 14     & 1.4     & 1.0     \\
           &        200 & 38     & 21     & 2.0     & 1.4     \\
           &        250 & 37     & 20     & 1.9     & 1.3     \\
\hline
AGN        &         15 & -          & 0.9     & 1.8     & 2.2     \\
           &         50 & -          & 1.8     & 0.7     & 0.4     \\
           &        100 & -          & 3.2     & 0.9     & 0.4     \\
           &        150 & -          & 6.1     & 1.4     & 0.7     \\
           &        200 & -          & 8.9     & 2.0     & 0.9     \\
           &        250 & -          & 8.8     & 1.9     & 0.8     \\
\hline
CNDN       &         15 & 1.0     & 1.4     & 0.8     & 1.4     \\
           &         50 & 3.0     & 2.9     & 0.3     & 0.3     \\
           &        100 & 5.5     & 5.1     & 0.4     & 0.3     \\
           &        150 & 10     & 9.7     & 0.6     & 0.4     \\
           &        200 & 16     & 14     & 0.9     & 0.6     \\
           &        250 & 15     & 14     & 0.8     & 0.5     \\
\hline
CNDS       &         15 & 0.7     & 0.7     & 0.6     & -          \\
           &         50 & 2.0     & 1.4     & 0.2     & -          \\
           &        100 & 3.8     & 2.4     & 0.3     & -          \\
           &        150 & 7.4     & 4.7     & 0.4     & -          \\
           &        200 & 11     & 6.9     & 0.6     & -          \\
           &        250 & 10     & 6.8     & 0.6     & -          \\
\hline\hline
Median      &   15      &  1.8  &  1.4    &  1.8   &     3 \\
            &   50      &  5.0  &  2.9    &  0.7   &   0.5 \\
            &   100     &  9.2  &  5.1    &  0.9   &   0.5 \\
            &   150     &  18  &  9.7    &  1.4   &   0.8 \\
            &   200     &  27  &  14    &    2   &   1.1 \\
            &   250     &  26  &  14    &  1.9   &   1.1 \\
\hline
\end{tabular}
\end{table}

\begin{table}
\caption{Total CS column densities ($\times 10^{14}$ cm$^{-2}$) in NGC 1068 for each of the sub-regions of interest in the SB ring calculated from individual level densities for various temperatures, assuming LTE, optically thin emission and that the source fills the beam. }
\label{table:N_SB}
\centering
\begin{tabular}{c c c c c}
\hline\hline
Sub-Region & T (K) & CS(2-1) & CS(3-2)& CS(7-6)\\
\hline
NSB-a      &         15 & 2.2     & 1.4     & 3.6     \\
           &         50 & 6.3     & 2.8     & 0.7    \\
           &        100 & 11     & 4.8     & 0.7     \\
           &        150 & 23     & 9.2     & 1.0     \\
           &        200 & 34     & 13     & 1.4     \\
           &        250 & 33     & 13     & 1.3     \\
\hline
NSB-b      &         15 & 0.9     & 0.6     & 0.6     \\
           &         50 & 2.3     & 1.1     & 0.1     \\
           &        100 & 4.4     & 1.9     & 0.1     \\
           &        150 & 8.6     & 3.7     & 0.2     \\
           &        200 & 12     & 5.4     & 0.2     \\
           &        250 & 12     & 5.3     & 0.2     \\
\hline
NSB-c      &         15 & 0.4     & 0.4     & 0.3     \\
           &         50 & 1.1     & 0.7     & 0.05     \\
           &        100 & 2.1     & 1.2     & 0.05     \\
           &        150 & 4.2     & 2.4     & 0.08     \\
           &        200 & 6.2     & 3.5     & 0.1     \\
           &        250 & 6.2     & 3.4     & 0.1     \\
\hline
SSB-a      &         15 & 1.6     & 1.0     & 0.9     \\
           &         50 & 4.7     & 2.1     & 0.2     \\
           &        100 & 8.7     & 3.7     & 0.2     \\
           &        150 & 16     & 7.0     & 0.3     \\
           &        200 & 25     & 10     & 0.4     \\
           &        250 & 24     & 10     & 0.3     \\
\hline
SSB-b      &         15 & 1.3     & 0.7     & 0.7     \\
           &         50 & 3.7     & 1.4     & 0.1     \\
           &        100 & 6.9     & 2.4     & 0.1     \\
           &        150 & 13     & 4.6     & 0.2     \\
           &        200 & 20     & 6.7     & 0.3     \\
           &        250 & 20     & 6.6     & 0.3     \\
\hline
SSB-c      &         15 & 1.2     & 0.6     & 0.5     \\
           &         50 & 3.3     & 1.1     & 0.08     \\
           &        100 & 6.2     & 1.9     & 0.09     \\
           &        150 & 12     & 3.7     & 0.1     \\
           &        200 & 18     & 5.4     & 0.2     \\
           &        250 & 17     & 5.4     & 0.2     \\
\hline\hline
Median      &   15      &  1.2  &  0.6    &  0.6 \\
            &   50      &  3.5  &  1.2    &  0.1 \\
            &   100     &  6.6  &  2.2    &  0.1 \\
            &   150     &  12  &  4.2    &  0.2 \\
            &   200     &  19  &  6.0    &  0.2 \\
            &   250     &  18  &    6    &  0.2 \\
\hline
\end{tabular}
\end{table}

\subsection{Rotation diagram}
\label{sec:rot}

Using the values of $N_u$ calculated under the optically thin assumption, we are also able to analyse the various sub-regions using the rotational diagram method.

A rotational diagram is a plot of the natural logarithm of column density over statistical weight of each transition against the energy in kelvin of the upper energy level, \mbox{$\ln{({N_u}/{g_u})}$ vs. $E (K)$.} If the assumption of LTE is correct the data can be fit by a straight line. The negative inverse slope of this line gives the rotational temperature, which is equivalent to the kinetic energy of the gas if it is in LTE and is a lower bound otherwise. The intercept of the line also corresponds to the natural log of the total column density of the species, $\ln{N}$ \citep[for further details see][]{GandL}.

\begin{figure*}
	\centering
	\includegraphics[width=17cm]{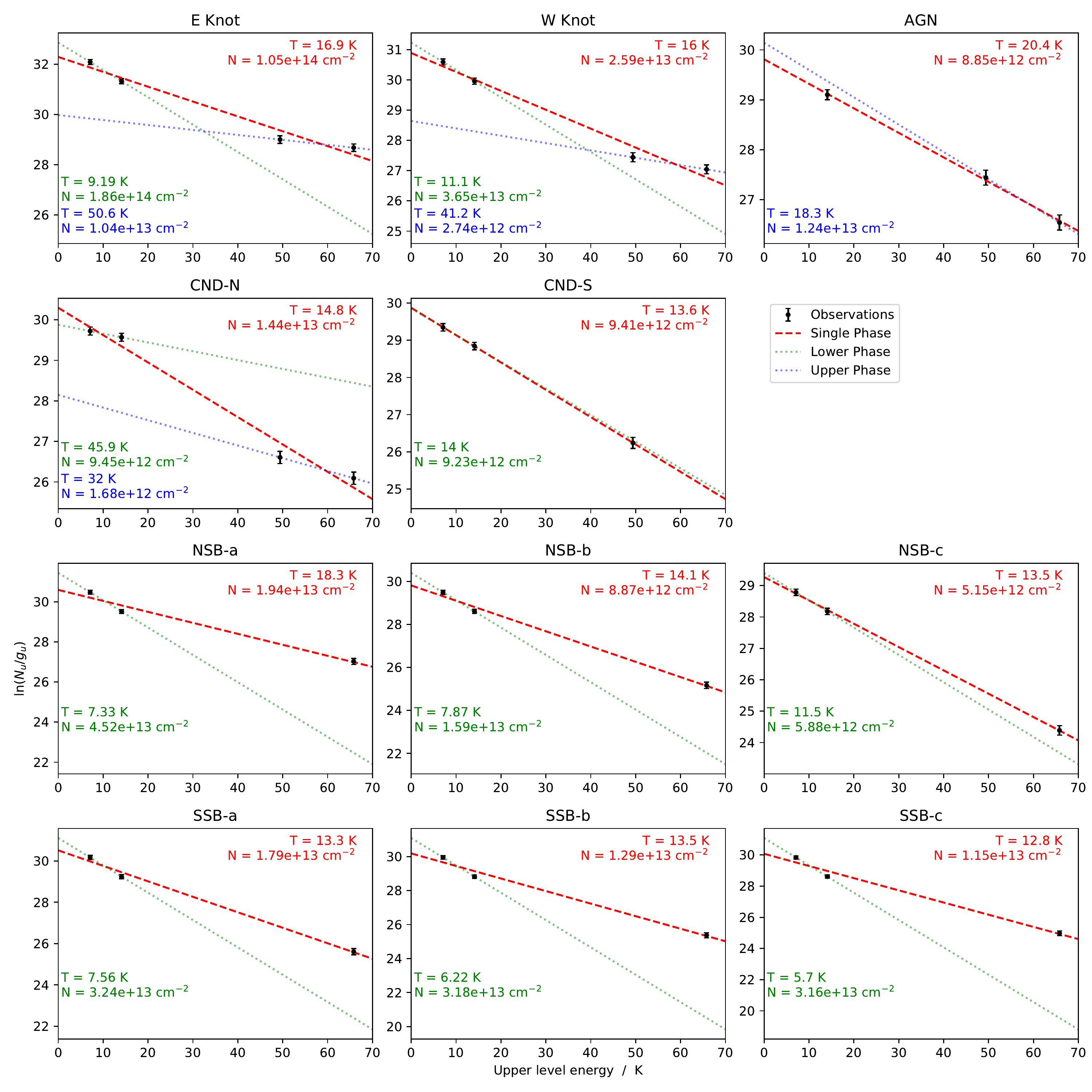}
	\caption{Rotational diagrams for CS in NGC 1068. The errors on the data are derived from the assumed error on the integrated intensities, 10-15\% depending on the band. The red dashed line shows the fit and derived conditions for all of the observed transitions. The green dotted line shows the fit and conditions for only the lower J transitions and the blue dotted line the same for the upper J.}
	\label{fig:rotDiag}
\end{figure*}

The rotation diagrams for the sub-regions of interest are shown in Fig. \ref{fig:rotDiag}. The error bars were obtained by assuming an error of 10\% on the transition column densities for transitions observed in band 3, and 15\% for bands 4 and 7 observations.

From the diagrams, we find the physical characteristics in each sub-region to be different, suggesting that conditions are not uniform across the galaxy, or indeed even across the CND. Looking at the CS column density, it appears largest in the E Knot at $\sim10^{14}$ cm$^{-2}$ and lowest in NSB-c at $5\times 10^{12}$ cm$^{-2}$. In the remaining sub-regions the column densities vary only slightly from $8\times 10^{12}$ cm$^{-2}$ to $2 \times 10^{13}$ cm$^{-2}$.

The rotational temperature appears to be greatest in the AGN and NSB-a, at $\sim$20 K and $\sim$18 K respectively, and lowest in the CND-S and remaining SB sub-regions, all of which have temperatures below $\sim$ 14 K. The temperatures of the remaining sub-regions are all in the range 15-17 K. Hence the inclusion of 15 K in the Boltzmann analysis (Tables \ref{table:N_CND} and \ref{table:N_SB}) to provide a comparison point to these results.

When comparing these results, the Boltzmann CS column densities are approximately an order of magnitude larger in all of the sub-regions. This is the same as found in \cite{Viti}, and may be due to assuming the gas to be optical thin. As can be seen in Equation 24 of \cite{GandL}, excluding the optical depth from calculation can lead to underestimating the ordinate in the rotational diagram, and hence also underestimating the total column densities. It could also be that the gas is not single phased, and that using a two phase fit may result in increased column densities.

\citet{Viti} found rotational temperatures of $\sim50$ K for the sub-regions in the CND looking at the CO 1-0, 2-1, 3-2 and 6-5 lines, and \citet{Garcia} found a dust temperature $T_{dust} \sim 46$ K from the dust emission, both of which are greater than the temperatures found here. Other works looking specifically at CS in the galaxy \citep{Nakajima15, Bayet3} give values of $\sim13$ K when looking at the CND as a whole, more in line with those seen here.


We have so far assumed that the gas is optically thin. In order to test that this assumption is correct, we now calculate the value of the opacity correction factor

\begin{equation}
	C_\tau = \frac{\tau}{1 - e^{-\tau}}
\end{equation}

\noindent where

\begin{equation}
	\tau = \frac{h}{\Delta\nu}N_uB_{ul}\left(e^{\frac{h\nu}{kT}}-1\right), \label{eqn:tau}
\end{equation}

\noindent and $B_{ul}$ is the Einstein B-coefficient. The values used for $T$ and $N_u$ are those obtained from the previous rotational diagrams. The value of $C_\tau$ obtained can then be used to calculate new values for $N_u$ for each transition using equation \ref{Nu}, and from these new rotational diagrams produced. This process is repeated iteratively until a stable value is found for $C_\tau$.

We immediately find $C_\tau \approx 1$ for each sub-region, such that the values calculated from our new rotational diagrams remain unchanged, and so we cease iterating. This suggests that the emission is optically thin, we are however aware of other studies which identify CS emission as optically thick \citep[e.g.][]{Linke}.


If the gas is optically thin, the apparent disagreement of the column densities between the Boltzmann analysis and rotational diagrams may instead indicate that the CS traces multiple phases of gas. Therefore, in regions where the required transitions were detected line fits were also obtained for the CS(2-1) and CS(3-2) transitions only, and similarly for CS(6-5) and CS(7-6) only. In sub-regions where all 4 transitions were detected the upper J fits give temperatures ($\sim$ 30-50 K) more in line with those observed for CO. The lower J fits gives higher CS column densities than the upper J in all of the sub-regions and temperatures ($\sim$ 5-15 K), lower than those obtained for single phase fits (other than in the CND-N, for which the temperature is higher than that given by the high J phase). These column densities are still lower than those found in Table \ref{table:N_CND}, and as such we are unsure as to the cause of the discrepancy between the two methods. Rotational diagram analysis of the galaxy as a whole by \citet{Bayet3} gives similar temperature results when using a two phase fits.

The phase traced by the low J transitions likely corresponds to a diffuse gas component. The reason for the low temperatures observed for this phase could then be explained by the gas being sub-thermal, too diffuse to be excited to higher temperatures. The phase traced by the higher J transitions on the other hand likely corresponds to dense cores. Such cores would not be large enough to fill the beam, suggesting that the assumption of a beam filling factor of unity is insufficient. We explore this further in section \ref{sec:nLTE}.


For each of the single phase models, the CS column densities obtained from the rotational analysis were compared to the average Boltzmann values (Tables \ref{table:N_CND} and \ref{table:N_SB}) calculated for that sub-region at the kinetic temperature closest to the obtained rotational temperature. The same was done for the values obtained using the lower and higher J transitions only, using the average of the relevant transitions. In all cases the Boltzmann values were larger by approximately an order of magnitude. The rotational and kinetic temperature are only equivalent if all levels are thermalised and optically thin. We have previously identified the gas as optically thin and so it is most likely the difference is due to the levels not all being thermalised, and hence the gas not in LTE.

Due to the low number of transitions in the rotational diagrams, we are unable to use these results to reliably constrain the conditions in each of the sub-regions. Nor can we differentiate between the gas being two phased or sub-thermally excited. We are, however, able to use them to rule out the optical depth as the cause for the discrepancy in column density calculated from each transition in Tables \ref{table:N_CND} and \ref{table:N_SB}. Instead, these results point to the gas not being in LTE.

\section{Non-LTE analysis}
\label{sec:nLTE}

\subsection{RADEX}
\label{sec:RADEX}

In order to fit the emission in each sub-region, the radiative transfer code RADEX \citep{RADEX} was used to model the observed CS spectral line energy distributions (SLEDs). Three different options are available for the applied escape probability method, that of a uniform sphere, an expanding sphere (LVG) or a plane parallel slab. Testing found that the geometry used had negligible impact on the results, and so the uniform sphere geometry was arbitrarily chosen.

The molecular data files used in the simulation were obtained from the LAMDA database \citep{LAMDA}, using H$_2$ as the collision partner. The background temperature in each sub-region was set equal to the cosmic background, 2.73 K, though this may not be completely valid for the AGN sub-region.

For each sub-region, the line width used was set to a constant value based on the spectra observed in Appendix \ref{sec:emission}. This was taken to be 150 km s$^{-1}$ within the CND and 50 km s$^{-1}$ within the SB ring, where the transitions appear narrower.

By varying the H$_2$ density, CS column density and kinetic  temperature in the model, a grid of $\sim 85~000$ points was produced over the following parameter space, based on the results of \citet{Viti} discussed previously.

\begin{itemize}
	\item H$_2$ density: $10^3 - 10^7$ cm$^{-3}$
	\item CS column density: $10^{12} - 10^{18}$ cm$^{-2}$
	\item Kinetic temperature: $10-200$ K
\end{itemize}

\begin{figure*}
\centering
   \includegraphics[width=17cm]{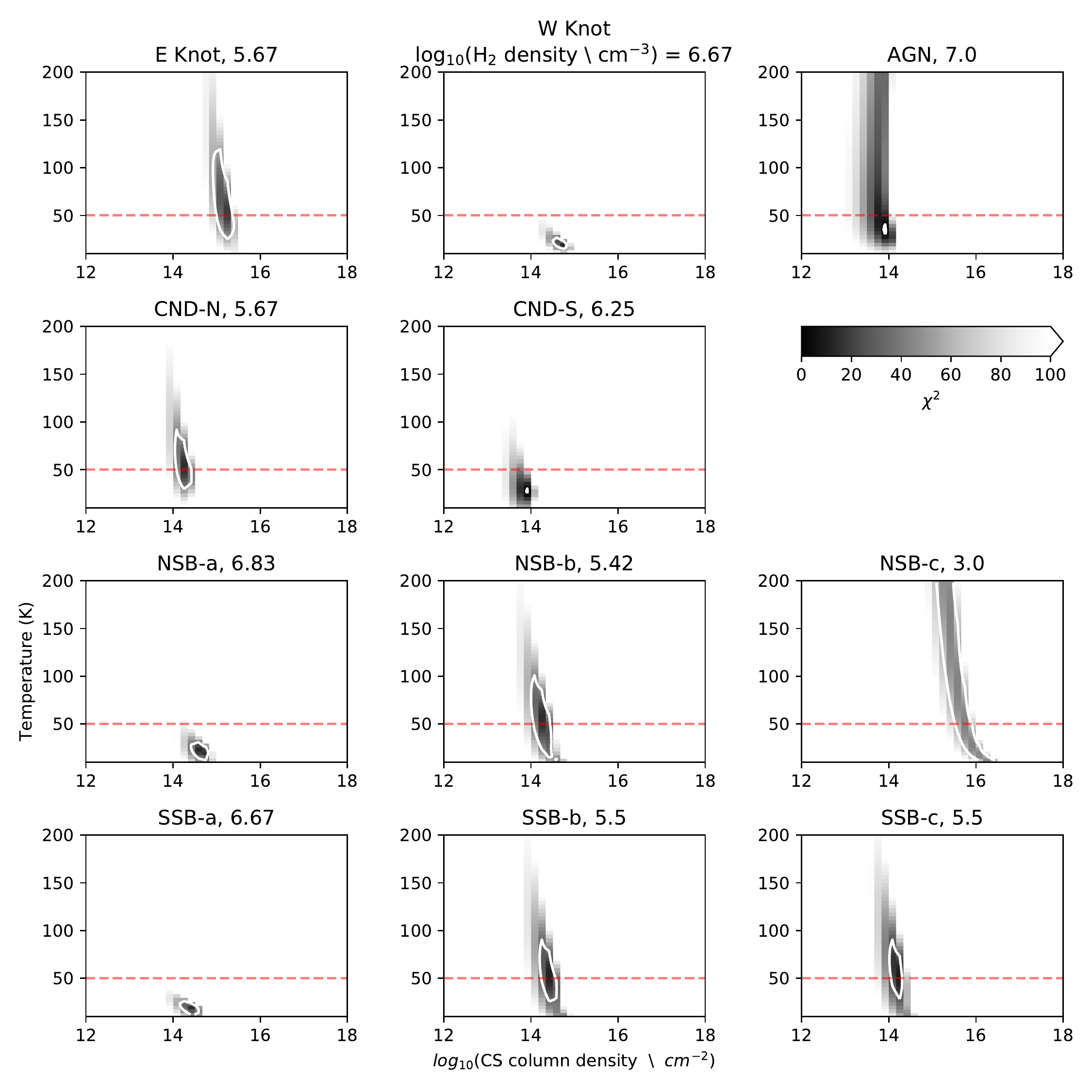}
     \caption{The $\chi^2$ fitting results for the RADEX modelling of the CS SLEDs in the various sub-regions of NGC 1068. Darker regions correspond to better fits indicated by lower $\chi^2$ values. The white contours show the range of $\chi^2$ used to determine the best fitting models, which varies between the sub-region. The NSB-c contour corresponds to $1.5\chi_{min}^2$ due to the large value of $\chi_{min}^2$, and similarly the AGN contour to a value of $3$ due to the small value of $\chi_{min}^2$. For the remaining sub-regions the contour is placed at $3\chi^2_{min}$. Slices of constant hydrogen density are shown, chosen to correspond to the lowest value of $\chi^2$. The exact value of H$_2$ density is given above each panel (as a decadic log).}
     \label{fig:chi_nH}
\end{figure*}

We found the upper bound of the temperature in each sub-region to be unconstrained within this temperature range. Testing a larger grid of temperatures with values up to $1,000$ K also resulted in unconstrained upper bounds for the temperatures. Thus we choose to use the CO rotational temperature of 50 K derived by \citet{Viti} as an upper bound for the temperatures when constraining the other parameters. This is similar to the dust temperature ($\sim46$ K) found in \cite{Garcia}.

The quality of the fits by the models was measured by comparing the main beam temperature predicted ($T_{mb}$) to the peak temperature ($T_{peak}$) observed using the reduced $\chi^2$ test. Slices of the resulting $\chi^2$ distribution cube at constant hydrogen density for all of the sub-regions are shown in Fig. \ref{fig:chi_nH}, and slices with alternative fixed variables in Appendix C. Different hydrogen densities are used for each sub-region, selected by taking the value which corresponds to the lowest $\chi^2$ for the sub-region.



For each sub-region, the contour corresponding to a value of $3\chi_{min}^2$ is plotted to give an indication of the range of parameters which best fit the observations. The NSB-c and AGN sub-regions are exceptions to this, with the contour instead placed at $1.5\chi_{min}^2$ and $3$ due to the large $(\sim 40)$ and small $(\sim 0.02)$ initial values of $\chi_{min}^2$ respectively. The values of CS column density and hydrogen density constrained from these fits are shown in Table \ref{tab:den}, as is the temperature of the best fitting model.



The constrained CS column densities are all approximately an order of magnitude greater than those predicted by the rotational diagram analysis, and hence similar to those predicted in the Boltzmann analysis. However in NSB-c, the sub-region in which the predicted column density is greatest, this difference is an order of magnitude greater when compared to both methods.


For the hydrogen densities, it was only possible to obtain a single limit for some of the sub-regions. Within the CND these sub-regions were the W Knot and the AGN. For both, we obtain lower limits. In the SB ring these sub-regions were NSB-a, SSB-a (both lower limits) and NSB-c (an upper limit). 

\begin{table*}
\caption{Summary of hydrogen and CS column densities from RADEX analysis of CS in each of the sub-regions in NGC 1068. Also shown is the temperature of the best fitting model, determined by the lowest $\chi^2$. The final column gives the value for the beam filling factor which best matches the model to the observations.}
\label{tab:den}
\centering
\begin{tabular}{c c c c c}
\hline\hline
Sub-region & log$_{10}$(n$_H$ [cm$^{-3}$]) & log$_{10}$(N$_{CS}$ [cm$^{-2}$]) & Temperature [K] & $\left(\frac{\Delta\Omega_a}{\Delta\Omega_s}\right)$ \\
\hline
E Knot 	& $5-6$	& $15-15.5$	& 50 & 0.95 \\
W Knot 	& $>6$	& $14.5-15$	& 18 & 0.94\\
AGN 	& $>6.5$ & $\sim14$	& 34 & 0.99	\\
CND-N 	& $5.5-6$	& $14-14.5$	& 50 & 1.00	\\
CND-S 	& $6-6.5$	& $\sim14$	& 26 & 0.95	\\
NSB-a 	& $>5.5$	& $14.5-15$	& 22 & 1.00	\\
NSB-b 	& $4.5-6$	& $14-14.5$	& 50 & 1.00	\\
NSB-c 	& $<4$	& $15-16$	& 50 & 1.00 \\
SSB-a 	& $>5$	& $14-14.5$	& 18 & 1.00	\\
SSB-b 	& $4.5-6$	& $14-14.5$	& 50 & 1.00 	\\
SSB-c 	& $5-6$	& $14-14.5$	& 50 & 0.92	\\
\hline
\end{tabular}
\end{table*}

The rotational diagram analysis (section \ref{sec:rot}) yields two gas components as traced by CS, a diffuse one traced by lower J transitions and one made up of dense cores traced by higher J lines. Such cores are too small to fill the beam, and so to account for this a beam filling factor which decreases with J should be used. However, as we cannot determine the size of the cores, we would need to adjust the beam filling factor as a free parameter in our model to find its correct value. The addition of this additional parameter would worsen the constraints we have for the other parameters, and so we forgo the inclusion in favour of our current fits.

Instead we identify the values a posteriori, using the difference between the model results and the observations. This was done by varying the beam filling factor assumed, and then comparing the corrected observations to the best fitting model previously selected for the sub-region, again using the $\chi^2$ test to evaluate the fits. The value which gives the best fit is then taken as the beam filling factor in that sub-region.

These obtained values are shown in table \ref{tab:den}, and are all $\sim1$. As it is unlikely that the gas completely fills the beam in all sub-regions these values suggest the conditions found to be incorrect, likely as a result of the models not considering the chemical feasibility of the parameters. As such, we next look at including a chemical code in our models.

\subsection{UCLCHEM}
\label{sec:UCL}

A limitation of the RADEX modelling is that it does not consider the chemical feasibility of the conditions used, such that the resulting best fit model may not be physical. In order to avoid this, the output of the chemical code UCLCHEM \citep{UCLCHEM} was coupled with RADEX in order to produce realistic parameters, and the resulting SLEDs for a grid of gas histories.

Modelling in UCLCHEM consists of two stages, a gas infall stage followed by a heating stage, starting with a cloud of diffuse atomic gas, at standard solar metallicity ISM abundances \citep[see][table 1]{abundances} and an initial density of 10 cm$^{-3}$.

The first stage allows the cloud to collapse by free-fall until it reaches a specified density, in order to create a high density cloud with self consistent abundances. This collapse is carried out at a fixed temperature of 10 K, and with the standard cosmic ray ionisation rate for molecular hydrogen of $\zeta_0 = 5\times 10^{-17} s^{-1}$ \citep{Williams}.

In the second stage the collapse is halted such that the density remains constant and a burst of star formation is simulated by heating the gas up to a specified temperature. The cloud is then allowed to evolve for $10^7$ years.

Three parameters were varied in the model, the density to which the cloud collapses during the first stage, the temperature it is heated to at the start of the second stage and the cosmic ray ionisation rate during the evolution in the second stage. The initial size of the cloud was also varied in tandem with the density, so as to ensure the resulting cloud had a visual extinction typical of its density (see respective values below).

A total of 36 models were run over the following parameter grid, chosen to cover a range of conditions:

\begin{itemize}
	\item Final density: $10^4, 10^5, 10^6$ cm$^{-3}$
	\begin{itemize}
		\item A$_v$: $10, 20, 50$ mag
	\end{itemize}
	\item Temperature: $50, 100, 300$ K
	\item $\zeta$: $1, 10, 10^2, 10^3~\zeta_0$
\end{itemize}


The motivation for these values are the same as in \citet{Viti}: the final densities and temperatures chosen are based on best fits found in this and previous studies of NGC 1068 and the large range of cosmic ray ionisation rates is intended to replicate the effect of an enhanced X-ray flux.

The output of UCLCHEM gives the visual extinction and final fractional abundances in the cloud. From these the CS column density is computed and input into RADEX alongside the temperature and density of the model to produce theoretical SLEDs for each of the sub-regions.

We are aware that sulphur is often considered to be depleted in chemical star formation models, however such considerations are beyond the scope of this work (for work on this topic see \citealt{Woods, Druard, Laas}).

\begin{figure*}
	\centering
		\includegraphics[width=17cm]{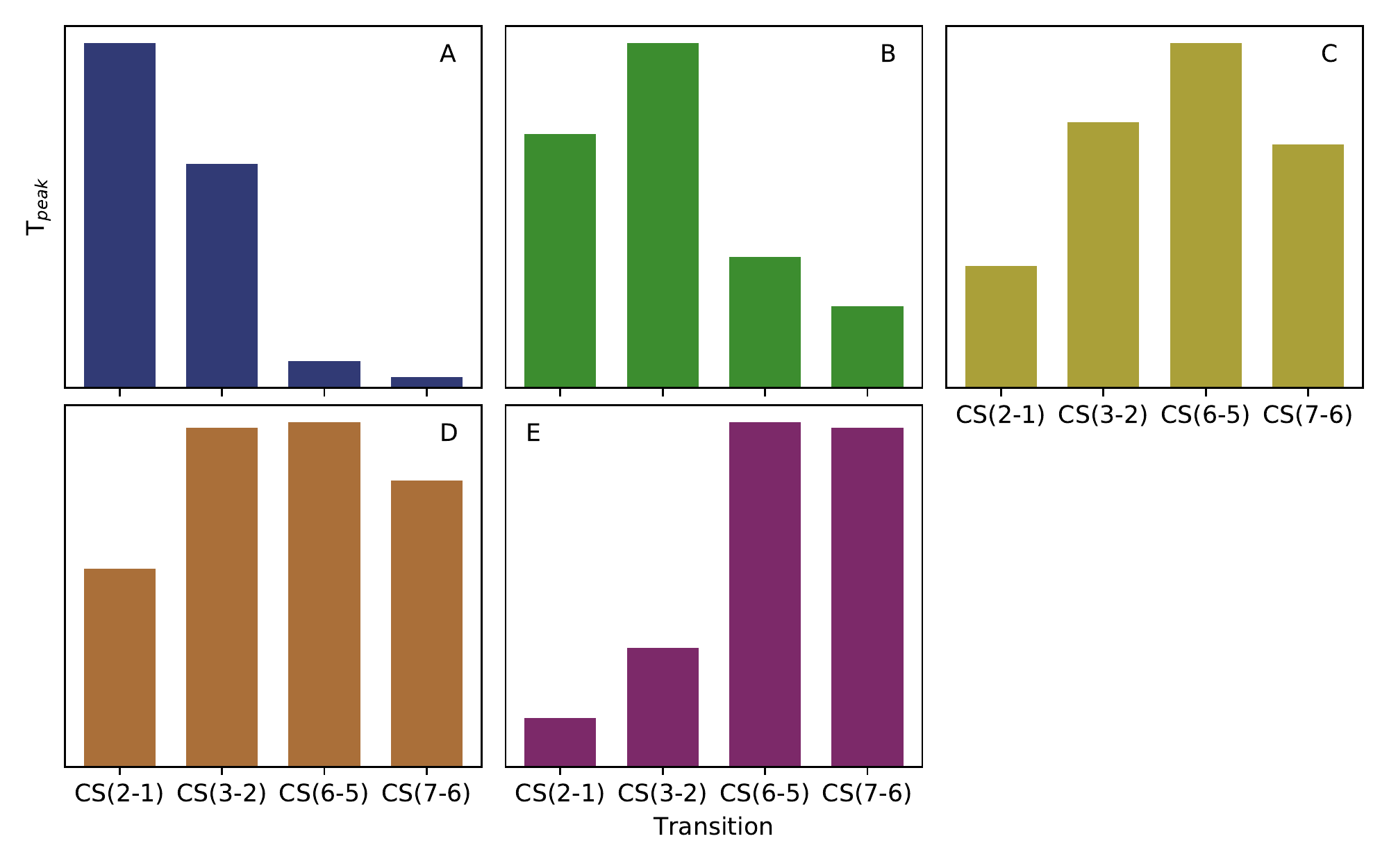}
			\caption{Examples of theoretical SLEDs obtained coupling UCLCHEM with RADEX in each of the groups. The absolute values of the predictions vary between the models in each group. Only the overall shape of the ladder is important in grouping and so these values are excluded.}
			\label{fig:modGroup}
\end{figure*}

To identify the best fitting model for each of the sub-regions, we match the shapes of the SLEDs from the models and observations by first splitting the former into 5 groups based on the shapes of their predicted ladders, independently of the physical parameters, labelled A through E. SLEDs in the first three groups are characterised by the transition for which their ladders peak at, CS(2-1), CS(3-2) and CS(6-5) for groups A, B and C respectively. In the remaining groups the two brightest transitions have approximately equal emission, such that there is no obvious peak in the ladders. These transitions are CS(3-2) and CS(6-5) in the case of group D, and CS(6-5) and CS(7-6) in the case of group E. Example SLEDs from each of these groups are shown in Fig. \ref{fig:modGroup}.

The general physical properties of models in each group are as follows. Models in group A have mid to low hydrogen densities, and cover all temperatures and cosmic ray ionisation rates. Group B includes models that have either high temperatures and medium hydrogen densities, or low temperatures and high hydrogen densities. Models in groups C and D both have middling temperatures and high hydrogen densities, with low cosmic ray ionisation rate models falling into group C, and high rates into group D. Finally, models in group E have high temperatures and hydrogen densities.


Due to the differing shape of the SLED in each of the sub-regions no single group characterises the emission in all sub-regions simultaneously. By extension this means no single model is able to match all of the sub-regions either, once again necessitating that each sub-region be fit independently.

\begin{figure*}
	\centering
		\includegraphics[width=17cm]{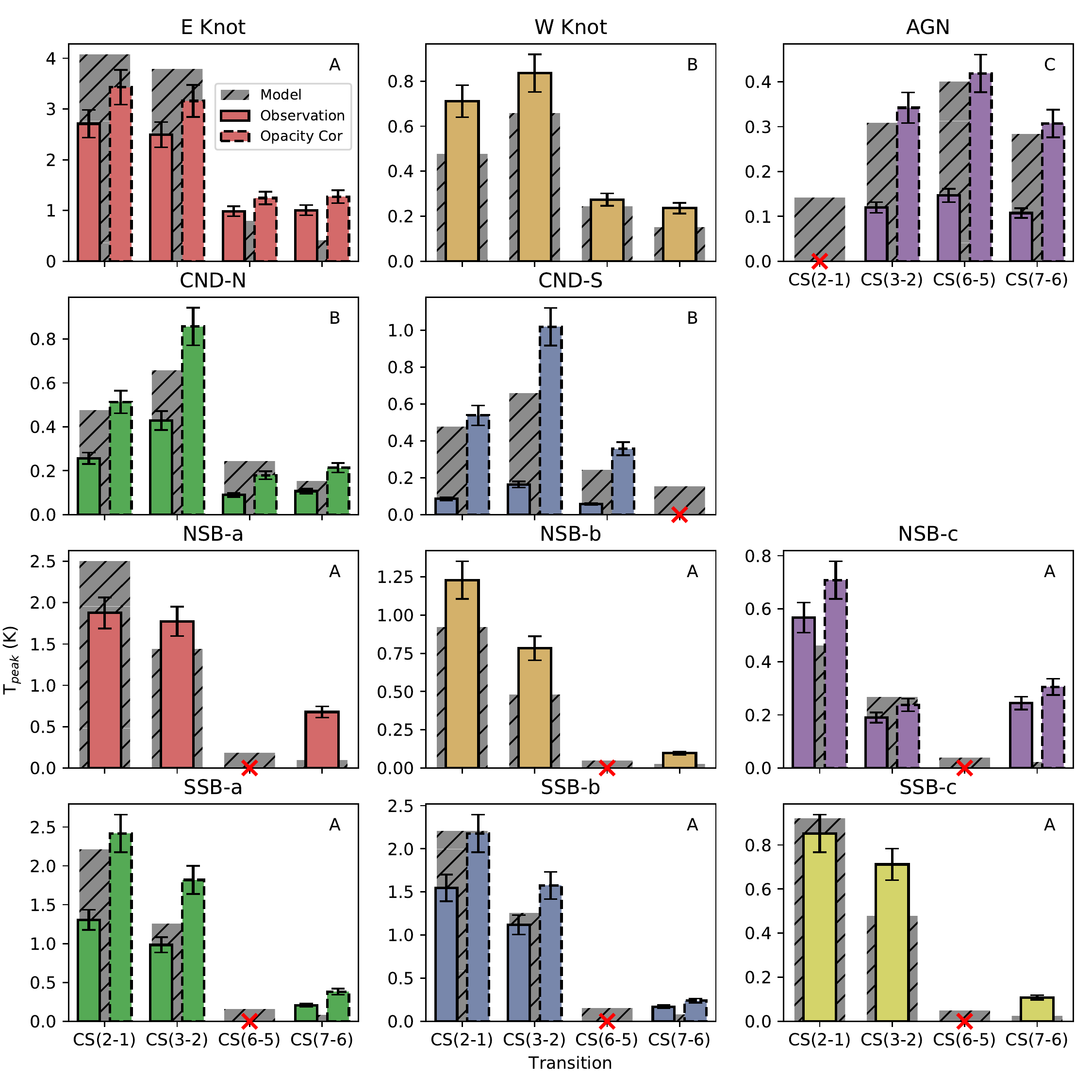}
			\caption{The observed SLED in each of the sub-regions, and the assigned model group. The grey hatched bars show the best fitting UCLCHEM \& RADEX model within that group. The coloured bars correspond to the observations, the solid outline shows the uncorrected observations and the dashed outline the observations corrected using the derived beam filling factor for the sub region. If this factor is found to be one, only a single bar is shown. Red Xs indicate transitions which were undetected in observations.}
			\label{fig:regGroup}
\end{figure*}

\begin{figure*}
\centering
	\includegraphics[width=17cm]{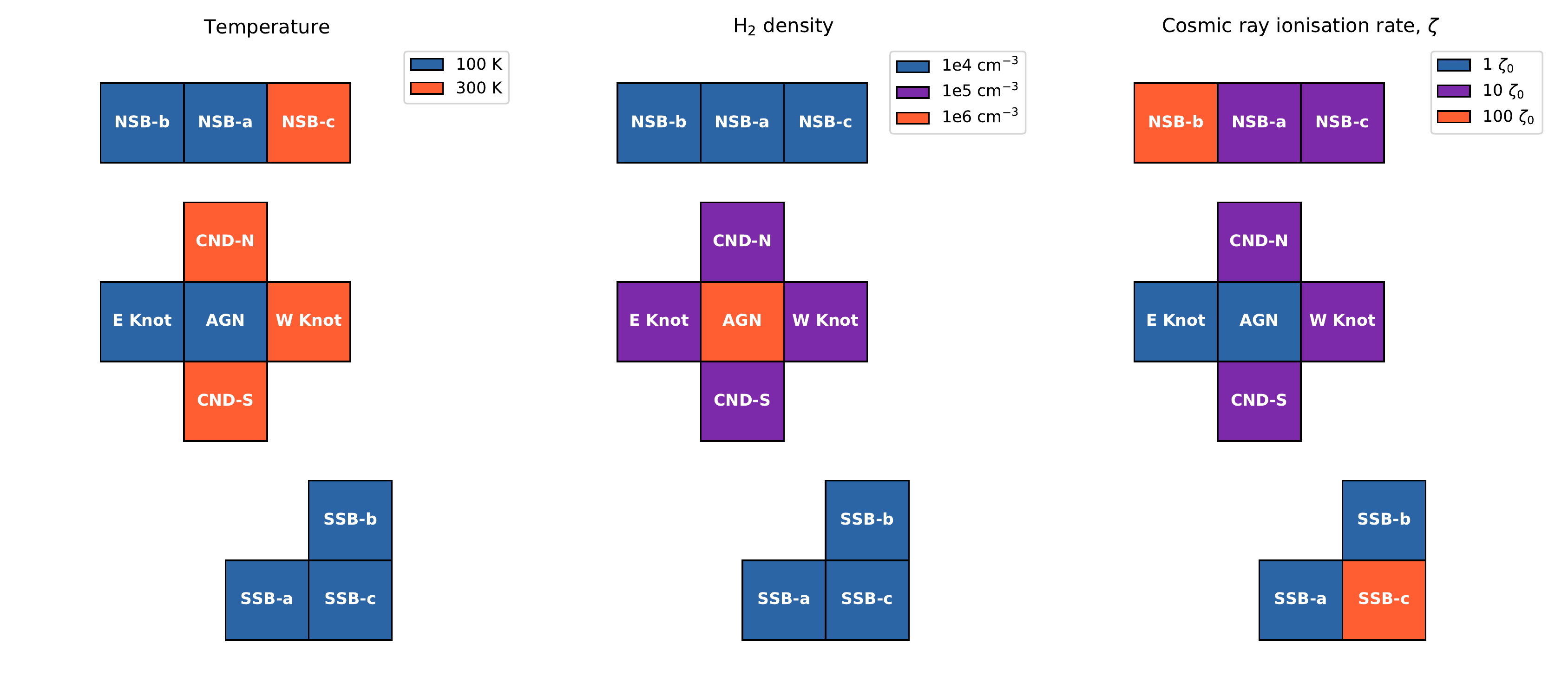}
		\caption{Cartoon depicting the best fit parameters from the UCLCHEM \& RADEX modelling} for each sub-region as in table \ref{table:UCL}.
		\label{fig:cartoon}
\end{figure*}

Using a standard $\chi^2$ test, the best fitting model from amongst those in the same group as the observations in each sub-region were selected. The observations and grouping of each sub-region and the corresponding best fitting model are shown in Fig. \ref{fig:regGroup}, and the parameters of the models in Table \ref{table:UCL}. Additionally, the parameters are presented in cartoon format in Fig. \ref{fig:cartoon}.

\begin{table*}
\caption{Summary of best fit temperature, hydrogen density, cosmic ray ionisation rate and CS column densities from the UCLCHEM \& RADEX modelling in each sub-region. Also shown are the CS fractional abundances and the beam filling factor which best matches the model to the observations.}
\label{table:UCL}      
\centering                                      
\begin{tabular}{c c c c c c c}          
\hline\hline                        
Region & Temperature (K) & log$_{10}$(n$_H$ [cm$^{-3}$]) & $\zeta$ [$\zeta_0$] & log$_{10}$(N$_{CS}$ [cm$^{-2}$]) & log$_{10}$(CS fraction) & $\left(\frac{\Delta\Omega_a}{\Delta\Omega_s}\right)$ \\    
\hline                                   
E Knot & 100 & 5 & 1 & 15.3 & -7.3 & 0.79 \\
W Knot & 300 & 5 & 10 & 14.3 & -8.2 & 1.0 \\
AGN & 100 & 6 & 1 & 14.1 & -8.8 & 0.35 \\
CND-N & 300 & 5 & 10 & 14.3 & -8.2 & 0.5 \\
CND-S & 300 & 5 & 10 & 14.3 & -8.2 & 0.16 \\
NSB-a & 100 & 4 & 10 & 15.7 & -6.6 & 1.0 \\
NSB-b & 100 & 4 & 100 & 15.1 & -7.1 & 1.0 \\
NSB-c & 300 & 4 & 10 & 14.5 & -7.7 & 0.8 \\
SSB-a & 100 & 4 & 1 & 15.6 & -6.6 & 0.54 \\
SSB-b & 100 & 4 & 1 & 15.6 & -6.6 & 0.71 \\
SSB-c & 100 & 4 & 100 & 15.1 & -7.1 & 1.0 \\
\hline
\end{tabular}
\end{table*}

Within the CND these results give the E Knot and AGN as the coolest sub-regions (both 100 K), suggesting an increase in temperature when moving from the east to the west of the region. The SB ring has similarly cool temperatures throughout with the exception of NSB-c, for which the temperature matches that found in the west of the CND (300 K).

Note that, due to the coarseness of the grid used, the actual values likely differ from those in the table. Using temperature as an example, there are no points between 50 K and 100 K and so the true temperature of the sub-regions in the SB ring are likely lower than the 100 K found through the modelling. It is also possible that the temperature is greater, however 100 K is already a high temperature for such gas so we assume this not to be the case.

The cosmic ray ionisation rate follows the same distribution as the temperature in the CND, being lower in the E Knot and AGN (1 $\zeta_0$) than in the other sub-regions (10 $\zeta_0$). The north SB arm has similar conditions to the west side of the CND and the south arm similar conditions to the east, with the exception of NSB-b and SSB-c, which have higher rates than any of the other sub-regions (100 $\zeta_0$). From the Pa$\alpha$ map (Fig. \ref{fig:alpha}), it appears that the rate of star formation in these sub-regions is greater than elsewhere in the ring, and so this is a likely cause for this increased rate.

Matching the LTE results, the hydrogen density in the CND is greater than that found in the SB ring across all sub-regions, with the AGN being the densest sub-region overall.

The CS fractional abundances (given here as decadic logs) seen in the CND match those commonly seen in SB galaxies \citep[$\sim -8.5$, see e.g.][Table 7]{Martin06}, with the exception of the E Knot which has higher abundances of $-7.3$, more in line with those seen in PDRs such as the the Orion Bar \citep[$-7.6$,][Table 9]{Martin06}, suggesting an abundance of massive stars in the sub-region.

Fractional abundances found in the SB region are mostly in the range $-7$ to $-6$, more than an order of magnitude greater than those seen in the CND and standard star forming regions. This is to be expected in these sub-regions as they are SB.

\begin{figure*}
	\centering
	   \includegraphics[width=17cm]{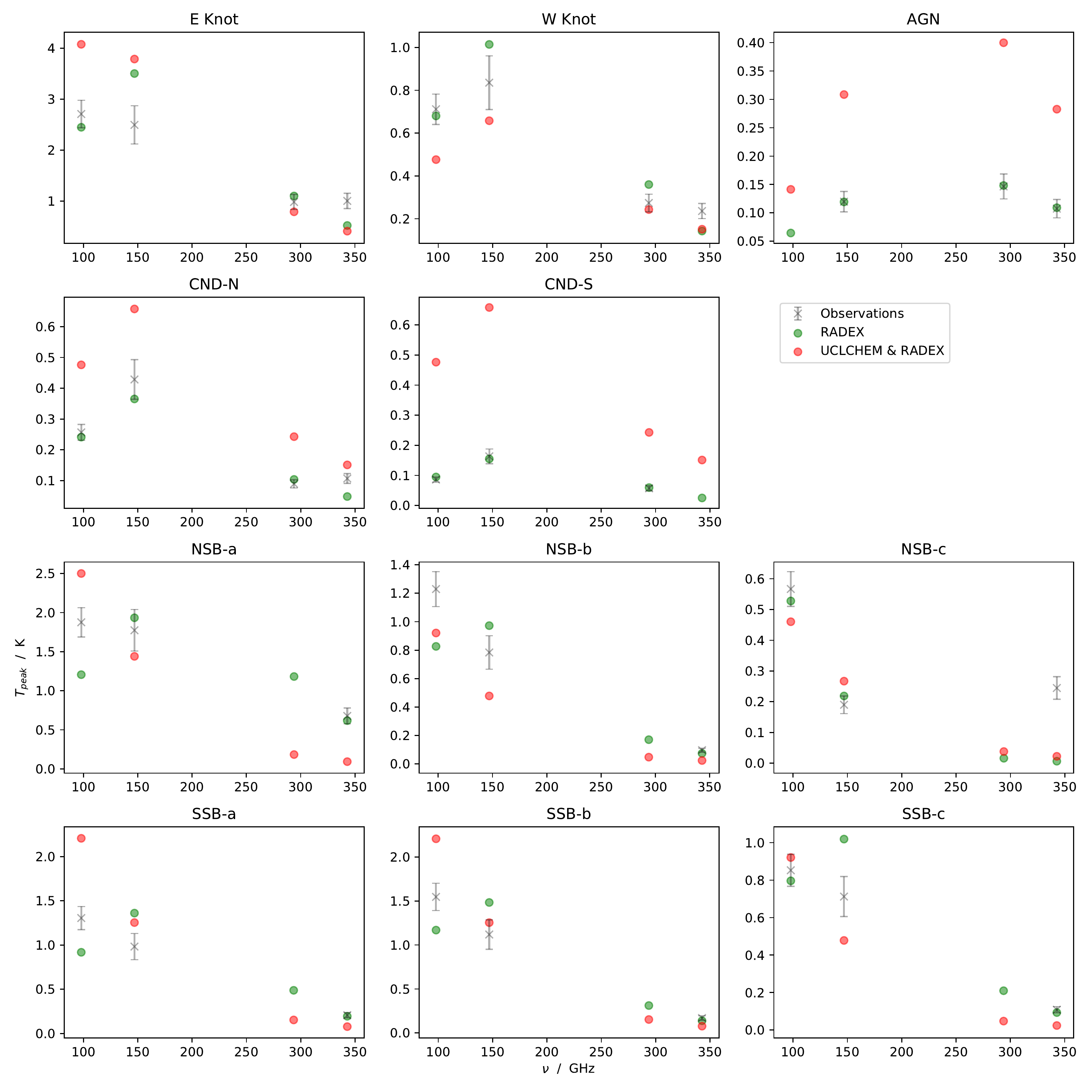}
		 \caption{Observed and theoretical CS spectral line energy distributions (SLEDs) for each sub-region. The points with error bars correspond to the observed SLEDs and the coloured points the SLED predicted by RADEX (green) and UCLCHEM coupled with RADEX (red). In the case that there was no detection for a transition, the modelled values are still shown. }
		\label{fig:ladder}
\end{figure*}

The SLEDs predicted for each sub-region by this model are shown in Fig. \ref{fig:ladder} alongside the actual observations. Also included in the figure are the SLEDs predicted by using RADEX along to model the ladders for comparison.

In many cases, it appears that the results from using RADEX alone give a better fit to the observations than the coupled models (e.g. for the AGN and CND-S sub-regions). This suggests that the CS column densities needed cannot be reproduced under the temperature and hydrogen densities as found by RADEX alone.

As discussed in section \ref{sec:RADEX}, it is unlikely that the beam is completely filled throughout our analysis, and so we attempt to calculate the value of the beam filling factor within each sub-region using the results of our UCLCHEM \& RADEX modelling to obtain a value a posteriori, as done in section \ref{sec:RADEX}. The values obtained are shown in table \ref{table:UCL}, with no obvious difference between those found in the CND and the SB ring regions. A comparison of the corrected observations to the models is included in Fig. \ref{fig:regGroup}.



One way in which these fits could be improved is to once again adopt a two phase model for the gas, as previous modelling suggests the transitions trace different gas. Accounting for the change in UV strength between the CND and the SB ring due to both the increased star formation rate in the latter, and its lower densities allowing further penetration could also improve the modelling.

It may also be necessary to include shocks in the modelling of the CND regions in order to accurately reproduce the observations. However as shocks enhance CS production this is unlikely to improve the fit in regions such as the AGN and CND-S, where the temperatures are already over predicted.

\section{Conclusions}
\label{sec:conc}

We present ALMA maps of the emission of the CS in the central $r\sim$ 2 kpc of NGC 1068, with spatial resolutions $\sim 0.4'' - 0.7''$ ($30-50$ pc) and perform a chemical analysis of the gas to identify the chemical differentiation between the CND and SB ring. As a result of our analysis we reach the following conclusions:


\begin{itemize}
	
	\item A common result across all of the modelling methods used is the requirement of different conditions to fit the emission observed in the various sub-regions, showing that the conditions in the galaxy are non-uniform.
	
	\item LTE analysis of the gas using Boltzmann's law shows the E Knot to be the densest sub-region in the CND, and the CND-S the least. This analysis also shows the SB ring being less dense than the CND.
	
	\item Rotation diagram analysis identifies the E Knot and NSB-c as the sub-regions of highest and lowest column densities in the galaxy respectively, and predicts rotational temperatures of $13-20$ K for all of the sub-regions with the AGN providing the highest value. The diagrams also give evidence for 2 gas phases, a diffuse and a dense phase.
	
	\item RADEX models were unable to constrain the temperature; however, by using the CO temperature found by \cite{Viti} as an upper limit, the hydrogen densities and CS column densities were constrained.
	
	\item Modelling of the gas using UCLCHEM and RADEX suggests that the temperature in the CND increases from east to west, and that the overall temperature is greater than in the SB ring. The cosmic ray ionisation rate follows a similar distribution to the temperature in the CND, with cooler sub-regions having lower rates. The  modelling also indicates that the CND ($10^5$ cm$^{-3}$) is denser than the SB ring ($10^4$ cm$^{-3}$), and the AGN is the densest sub-region overall ($10^6$ cm$^{-3}$).
	
		
\end{itemize}

RADEX modelling alone produces better matches to observations than RADEX/UCLCHEM coupled models (Fig. \ref{fig:ladder}). This suggests that, with the chemistry used for the UCLCHEM model, the CS column densities required by RADEX cannot be reproduced at the required temperatures and hydrogen densities. This difference is most pronounced in the AGN and CND-S sub-regions, both of which are identified as being characterized by a high fraction of hot (shocked) gas in \citet{Garcia}, implying the necessity of shock chemistry to reproduce the observations.

Analysis also indicates that the dense molecular gas is multi-phased, with the low and high J transitions tracing a diffuse gas and dense cores respectively. In order to further test this requires high-resolution CS transition maps covering both the CND and the SB ring for additional transitions. Such observations would also allow us to further constrain the physical parameters in the sub-regions by the various models used, as currently only 3-4 transitions are available for each which limits the reliability of our modelling efforts.

\section*{Acknowledgements}

This paper makes use of the following ALMA data: ADS/JAO.ALMA\#2011.0.00083.S, ADS/JAO.ALMA\#2013.1.00055.S, ADS/JAO.ALMA\#2015.1.01144.S.  ALMA is a partnership of ESO (representing its member states), NSF (USA) and NINS (Japan), together with NRC (Canada), MOST and ASIAA (Taiwan), and KASI (Republic of Korea), in cooperation with the Republic of Chile. The Joint ALMA Observatory is operated by ESO, AUI/NRAO and NAOJ.

SGB, AAH, and AU acknowledge support from grant PGC2018-094671-B-I00 (MCIU/AEI/FEDER, UE). AAH work was done under project  No. MDM-2017-0737 Unidad de Excelencia "Mar\'{\i}a de Maeztu"- Centro de Astrobiolog\'{\i}a (INTA-CSIC).

The authors thank the anonymous referee for constructive comments that improved the original version of the paper.

\section*{Data Availability}

The ALMA data used in this work is publicly available online via the ALMA archive, \url{http://almascience.nrao.edu/aq/}, with project codes are ADS/JAO.ALMA\#2011.0.00083.S, ADS/JAO.ALMA\#2013.1.00055.S, ADS/JAO.ALMA\#2015.1.01144.S. The maps produced from these data will be shared on reasonable request to the corresponding author.




\bibliographystyle{mnras}
\bibliography{ALMA_CS_NGC_1068_MNRAS} 



\appendix

\section{Emission maps and spectra}
\label{sec:emission}

Here we include the CS observations analysed in this work. In these figures the upper panel shows the velocity integrated intensity map, with beam size indicated by a hatched ellipse in the lower right. The remaining panels show the emission line profile of the transition in the different sub-regions between $\pm 250$ km/s, with the blue line indicating the fit to the emission line in the case that one was found and the red the $3\sigma$-rms detection threshold. Note that although some sub-regions appear to fall in areas of no detection for this transition, we find signals from other lines in those regions.

\subsection{SB ring observations}
\label{sec:SBemission}

The maps here cover the SB ring of NGC 1068 over the FOV of the observations, with spectra from sub-regions within the ring shown below. The CS(2-1) transition is shown in Fig. \ref{fig:cs21SB}, CS(3-2) in Fig. \ref{fig:cs32SB} and CS(7-6) in Fig. \ref{fig:cs76SB}. The CS(6-5) transition is not included as the 21$''$ FOV is insufficient to cover the SB ring.

\begin{figure*}
	\centering
	\includegraphics[width=15.4cm]{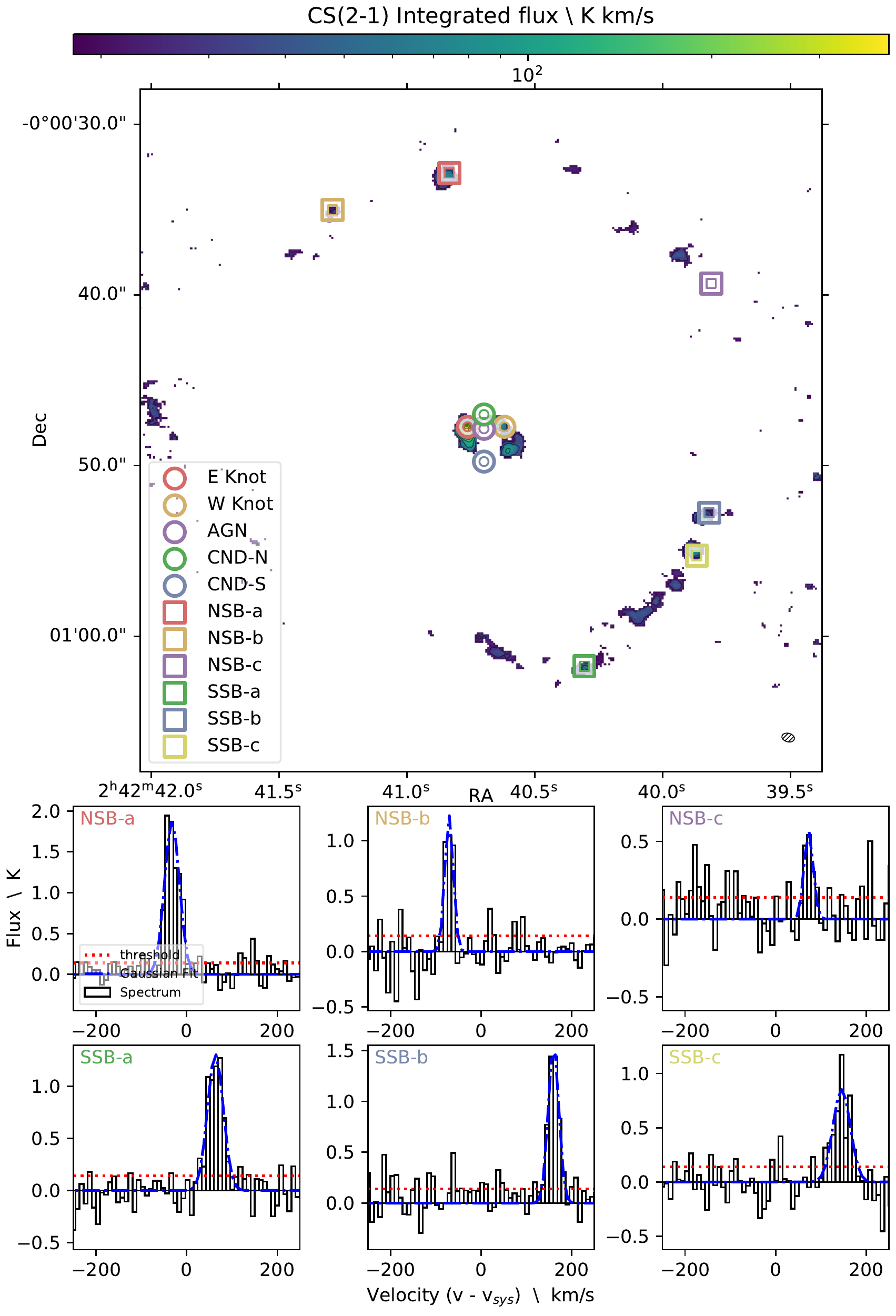}
    \caption{\small{CS(2-1) observations in the SB ring of NGC 1068. The upper panel shows the velocity integrated intensity map over the entire galaxy truncated to the 70$''$ FOV of observation. The colour map shows emission above a 2$\sigma$ threshold, and the contour levels are at $3\sigma$ and $5\sigma$ to $40\sigma$ in intervals of $5\sigma$, where the sigma is 9 K km s$^{-1}$.  The beam size $0.75'' \times 0.51''$ is indicated by the hatched ellipse in the lower right. The remaining panels show the emission line profile of the transition in the different sub-regions in the SB ring. The blue line shows the fit to the emission line in the case that one was found and the red the $3\sigma$-rms detection threshold of 0.42 K. Note that although some sub-regions appear to fall in areas of no detection for this transition, we find signals from other lines in those regions.}}
	\label{fig:cs21SB}
\end{figure*}

\begin{figure*}
	\centering
	\includegraphics[width=15.4cm]{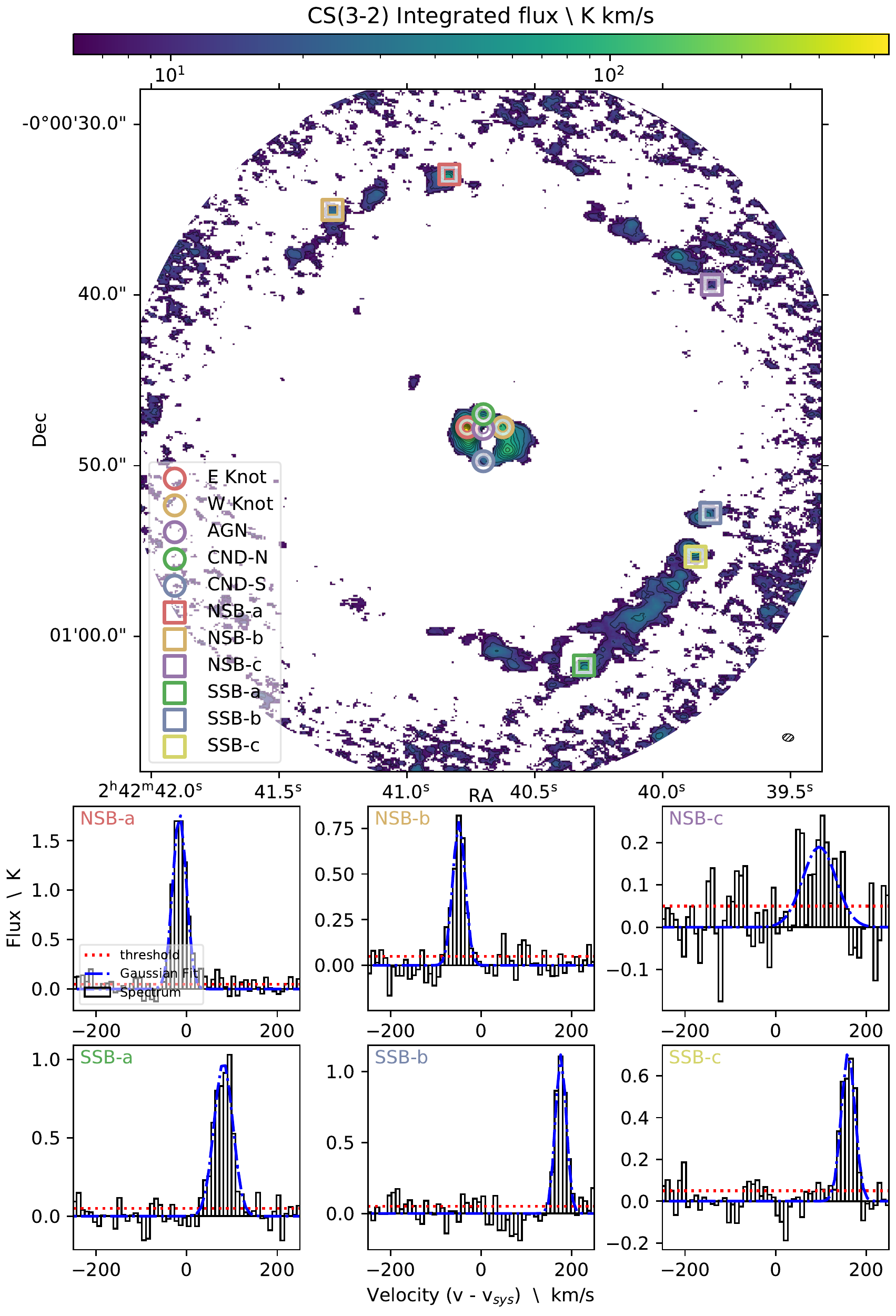}
	\caption{\small{CS(3-2) observations in the SB ring of NGC 1068. The upper panel shows the velocity integrated intensity map over the entire galaxy truncated to the 42$''$ FOV of observation. The colour map shows emission above a 2$\sigma$ threshold, and the contour levels are at $3\sigma$, $5\sigma$, $10\sigma$, $15\sigma$, $20\sigma$, $25\sigma$, $30\sigma$ and $35\sigma$ to $105\sigma$ in intervals of $10\sigma$, where the sigma is 3 K km s$^{-1}$. The beam size $0.66'' \times 0.44''$ is indicated by the hatched ellipse in the lower right. The remaining panels show the emission line profile of the transition in the different sub-regions in the SB ring. The blue line shows the fit to the emission line in the case that one was found and the red the $3\sigma$-rms detection threshold of 0.15 K.}}
	\label{fig:cs32SB}
\end{figure*}

\begin{figure*}
	\centering
	\includegraphics[width=15.4cm]{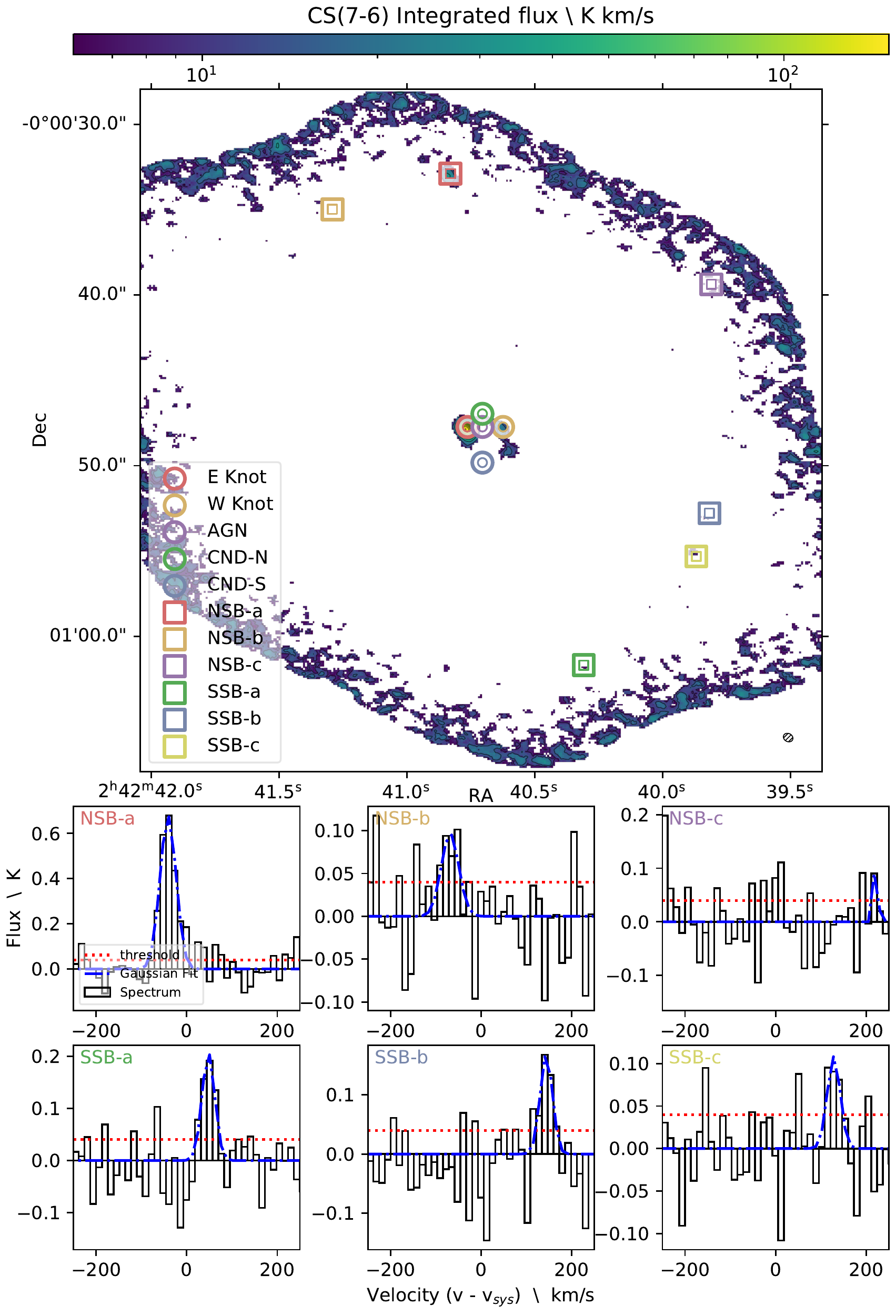}
	\caption{\small{CS(7-6) observations in the SB ring of NGC 1068. The upper panel shows the velocity integrated intensity map over the entire galaxy truncated to the 50$''$ FOV of observation. The colour map shows emission above a 2$\sigma$ threshold, and the contour levels are at $3\sigma$ and $5\sigma$ to $50\sigma$ in intervals of $5\sigma$, where the sigma is 3 K km s$^{-1}$. The beam size $0.59'' \times 0.48''$ is indicated by the hatched ellipse in the lower right. The remaining panels show the emission line profile of the transition in different sub-regions in the SB ring. The blue line shows the fit to the emission line in the case that one was found and the red the $3\sigma$-rms detection threshold of 0.12 K. Note that although some sub-regions appear to fall in areas of no detection for this transition, we find signals from other lines in those regions.}}
	\label{fig:cs76SB}
\end{figure*}

\FloatBarrier
\subsection{CND ring observations}
\label{sec:CNDemission}

\begin{figure*}
	\centering
	\includegraphics[width=15.4cm]{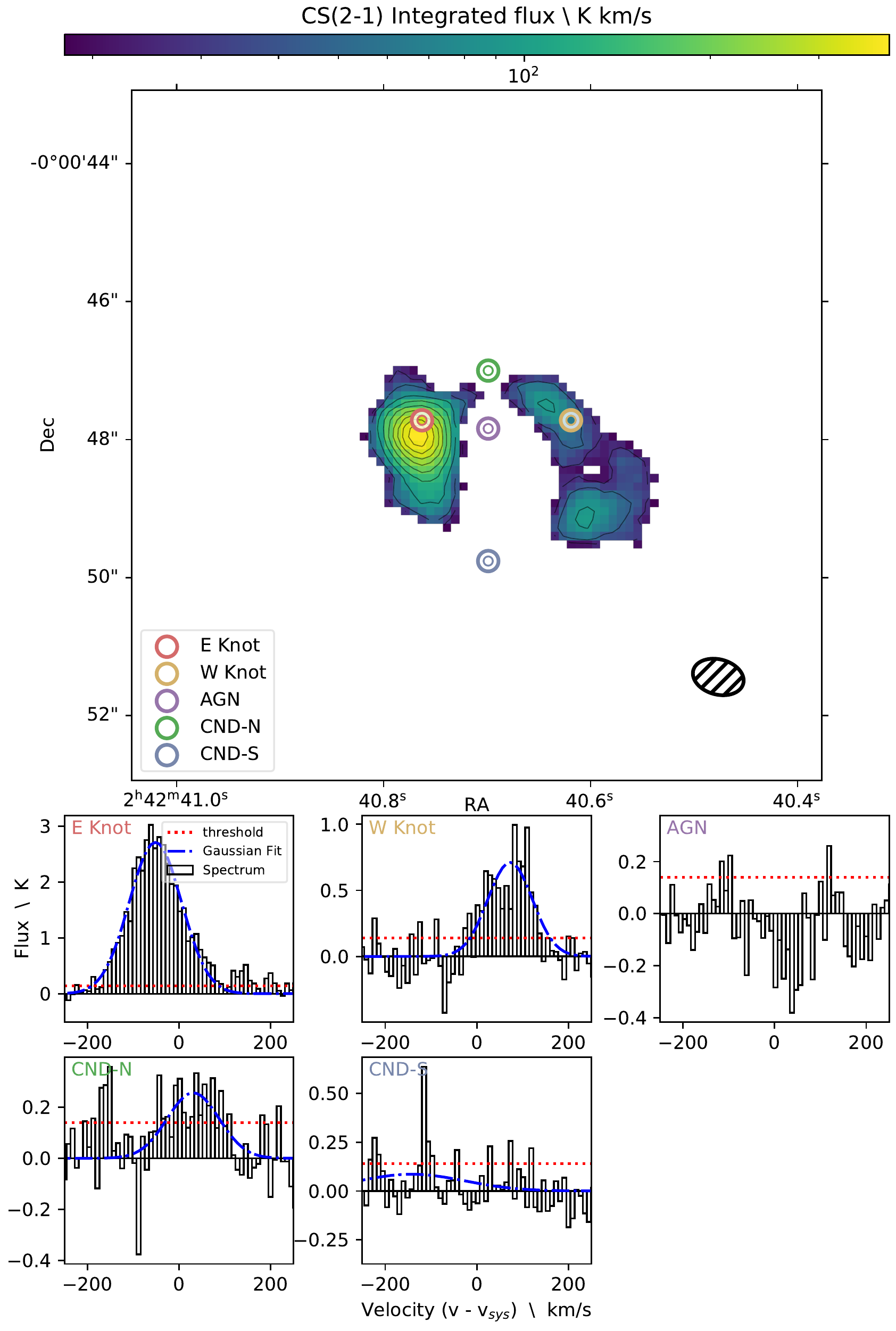}
	\caption{\small{CS(2-1) observations in the CND of NGC 1068. The upper panel shows the velocity integrated intensity map over the CND. The colour map shows emission above a 2$\sigma$ threshold, and the contour levels are at $3\sigma$ and $5\sigma$ to $40\sigma$ in intervals of $5\sigma$, where the sigma is 9 K km s$^{-1}$. The beam size $0.75'' \times 0.51''$ is indicated by the hatched ellipse in the lower right. The remaining panels show the emission line profile of the transition in the different sub-regions in the CND. The blue line shows the fit to the emission line in the case that one was found and the red the $3\sigma$-rms detection threshold of 0.42 K. Note that although some sub-regions appear to fall in areas of no detection for this transition, we find signals from other lines in those regions.}}
	\label{fig:cs21CND}
\end{figure*}

\begin{figure*}
	\centering
	\includegraphics[width=15.4cm]{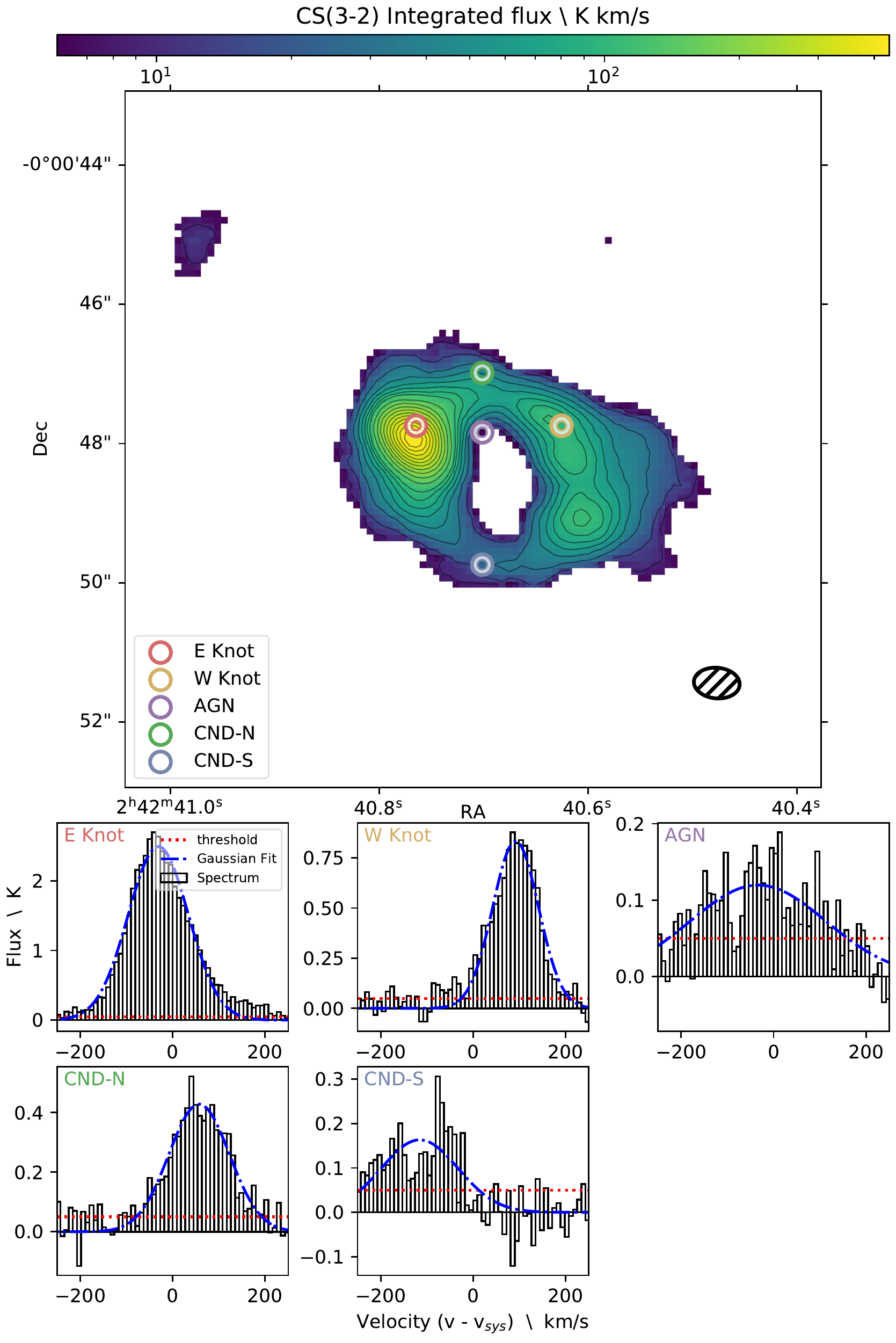}
	\caption{\small{CS(3-2) observations in the CND of NGC 1068. The upper panel shows the velocity integrated intensity map over the CND. The colour map shows emission above a 2$\sigma$ threshold, and the contour levels are at $3\sigma$, $5\sigma$, $10\sigma$, $15\sigma$, $20\sigma$, $25\sigma$, $30\sigma$ and $35\sigma$ to $105\sigma$ in intervals of $10\sigma$, where the sigma is 3 K km s$^{-1}$. The beam size $0.66'' \times 0.44''$ is indicated by the hatched ellipse in the lower right. The remaining panels show the emission line profile of the transition in the different sub-regions in the CND. The blue line shows the fit to the emission line in the case that one was found and the red the $3\sigma$-rms detection threshold of 0.15 K.}}
	\label{fig:cs32CND}
\end{figure*}

\begin{figure*}
	\centering
	\includegraphics[width=15.4cm]{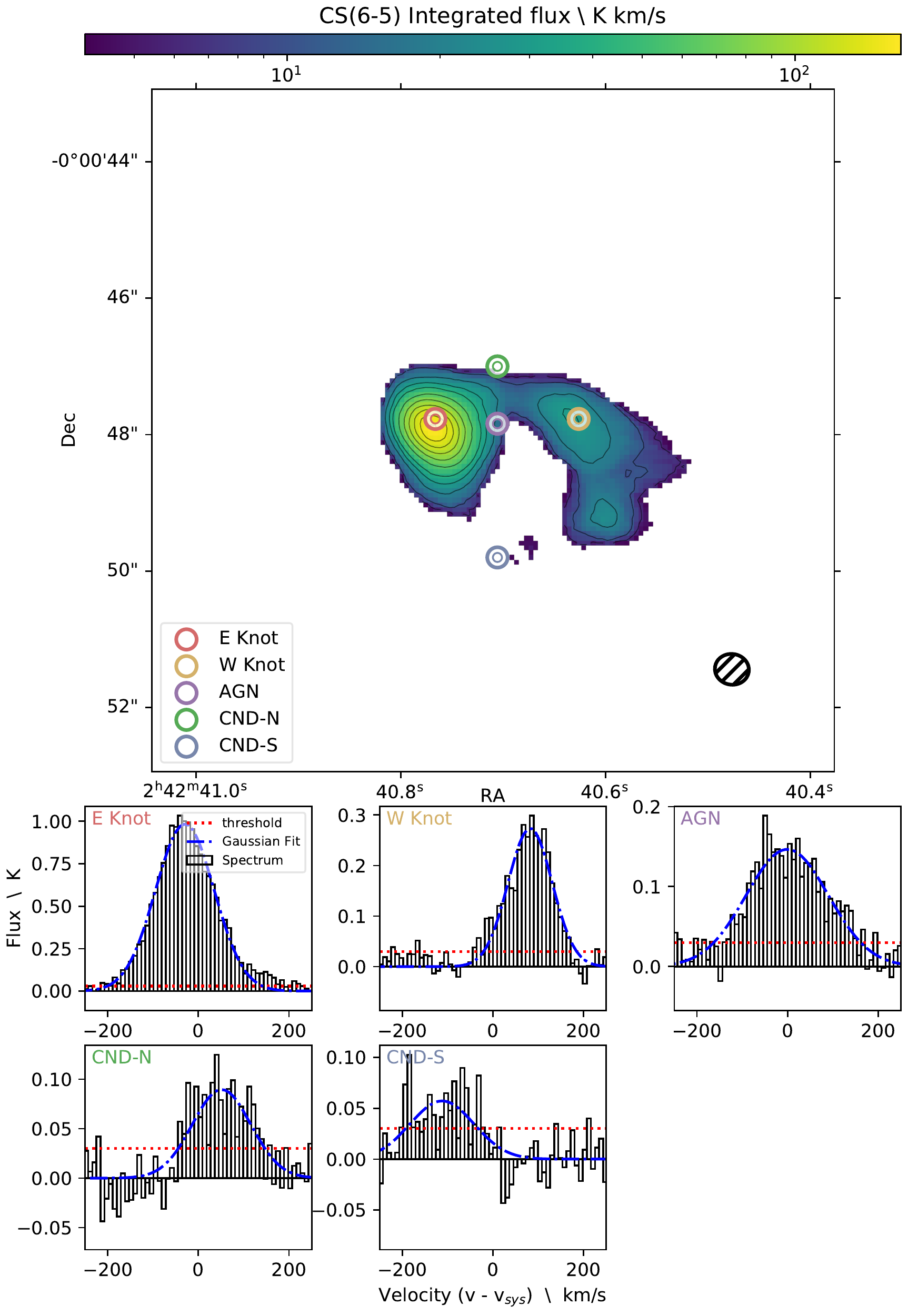}
	\caption{\small{CS(6-5) observations in the CND of NGC 1068. The upper panel shows the velocity integrated intensity map over the CND. The colour map shows emission above a 2$\sigma$ threshold, and the contour levels are at $3\sigma$, $5\sigma$, $10\sigma$, $15\sigma$ and $20\sigma$ to $80\sigma$ in intervals of $10\sigma$, where the sigma is 2 K km s$^{-1}$. The beam size $0.50'' \times 0.45''$ is indicated by the hatched ellipse in the lower right. The remaining panels show the emission line profile of the transition in the different sub-regions in the CND. The blue line shows the fit to the emission line in the case that one was found and the red the $3\sigma$-rms detection threshold of 0.09 K. Note that although some sub-regions appear to fall in areas of no detection for this transition, we find signals from other lines in those regions.}} 
	\label{fig:cs65CND}
\end{figure*}

\begin{figure*}
	\centering
	\includegraphics[width=15.4cm]{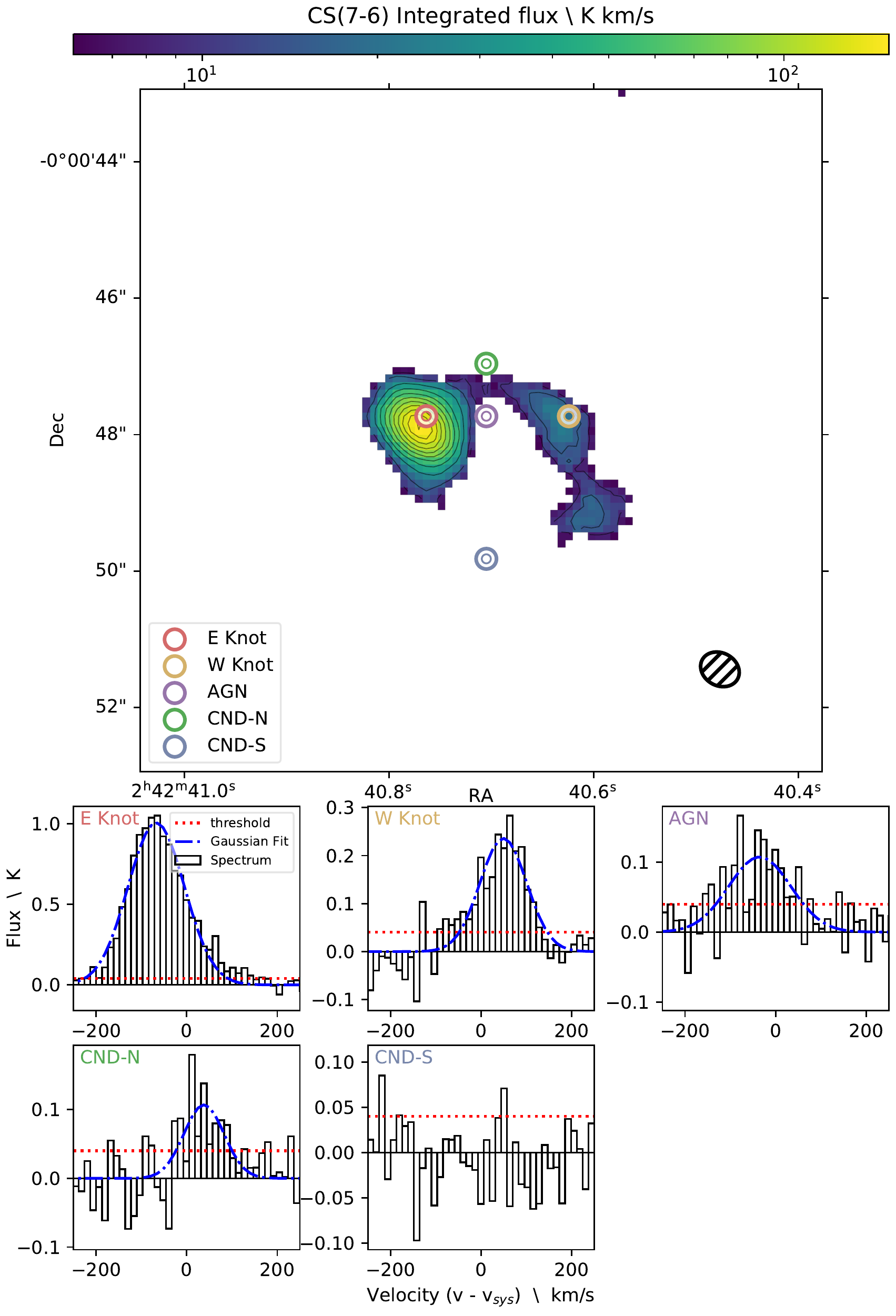}
	\caption{\small{CS(7-6) observations in the CND of NGC 1068. The upper panel shows the velocity integrated intensity map over the CND. The colour map shows emission above a 2$\sigma$ threshold, and the contour levels are at $3\sigma$ and $5\sigma$ to $50\sigma$ in intervals of $5\sigma$, where the sigma is 3 K km s$^{-1}$. The beam size $0.59'' \times 0.48''$ is indicated by the hatched ellipse in the lower right. The remaining panels show the emission line profile of the transition in different sub-regions in the CND. The blue line shows the fit to the emission line in the case that one was found and the red the $3\sigma$-rms detection threshold of 0.12 K. Note that although some sub-regions appear to fall in areas of no detection for this transition, we find signals from other lines in those regions.}}
	\label{fig:cs76CND}
\end{figure*}

These maps are the same as those in \ref{sec:SBemission} but zoomed in on the CND region, with spectra from sub-regions within shown. The CS(2-1), CS(3-2), CS(6-5) and CS(7-6) transitions are shown in Figs. \ref{fig:cs21CND}, \ref{fig:cs32CND}, \ref{fig:cs65CND} and \ref{fig:cs76CND} respectively

\FloatBarrier
\section{Ratio maps}
\label{sec:ratio}

\renewcommand{\thefigure}{B\arabic{figure}}
\setcounter{figure}{0}

\begin{figure*}
	\centering
    \includegraphics[width=11cm]{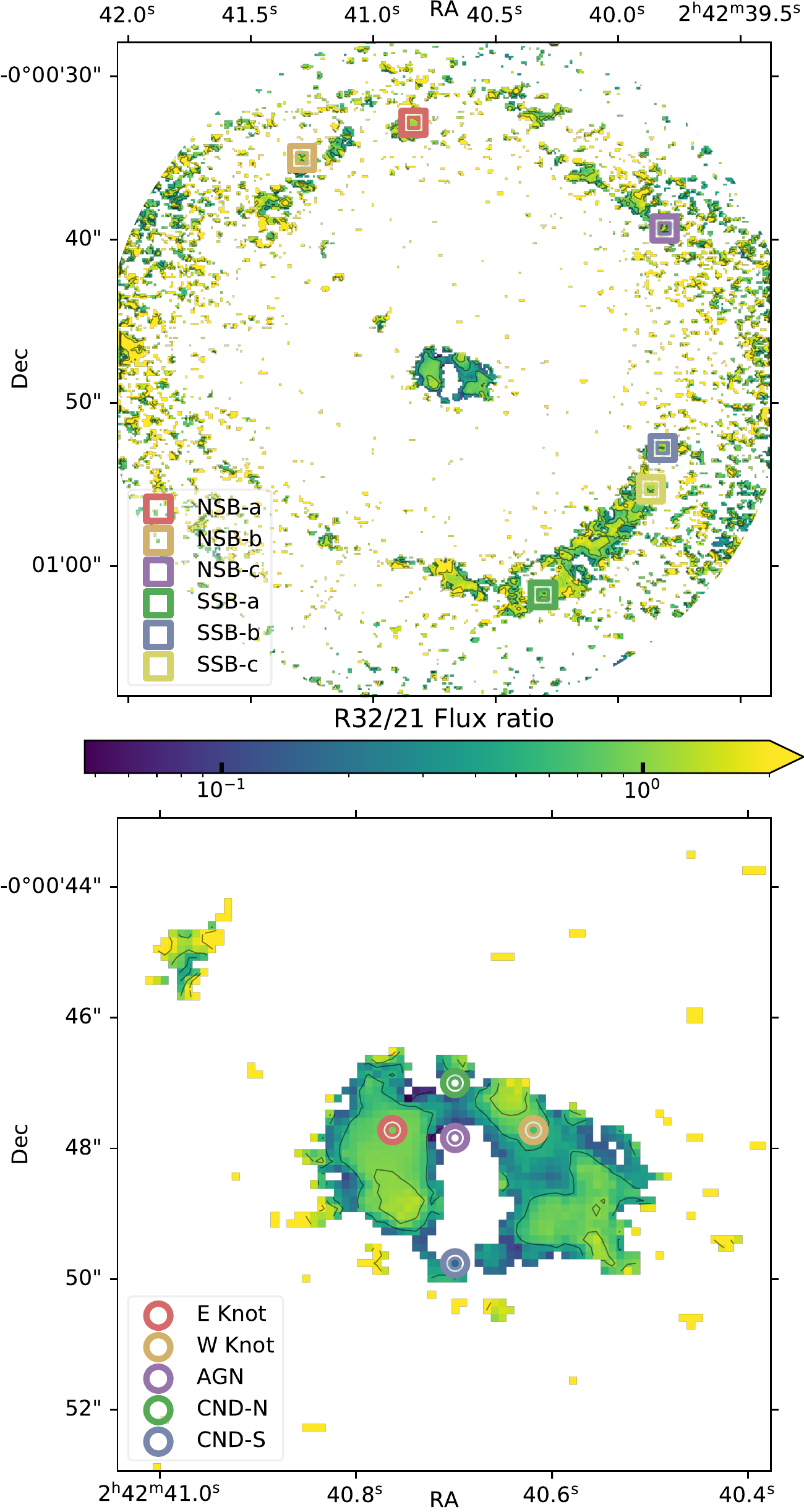}
	\caption{The line ratio CS(3-2)/CS(2-1) over the SB ring \textit{(top)} and the CND \textit{(bottom)}. The map was produced by degrading the CS(3-2) map to the same resolution as the CS(2-1) map ($0.75'' \times 0.51''$) and is truncated at the 42$''$ FOV of the CS(3-2) map. The contours correspond to 0.1, 0.5, 1 and 2.}
	\label{fig:r32/21}
\end{figure*}

\begin{figure*}
	\centering
    \includegraphics[width=12.5cm]{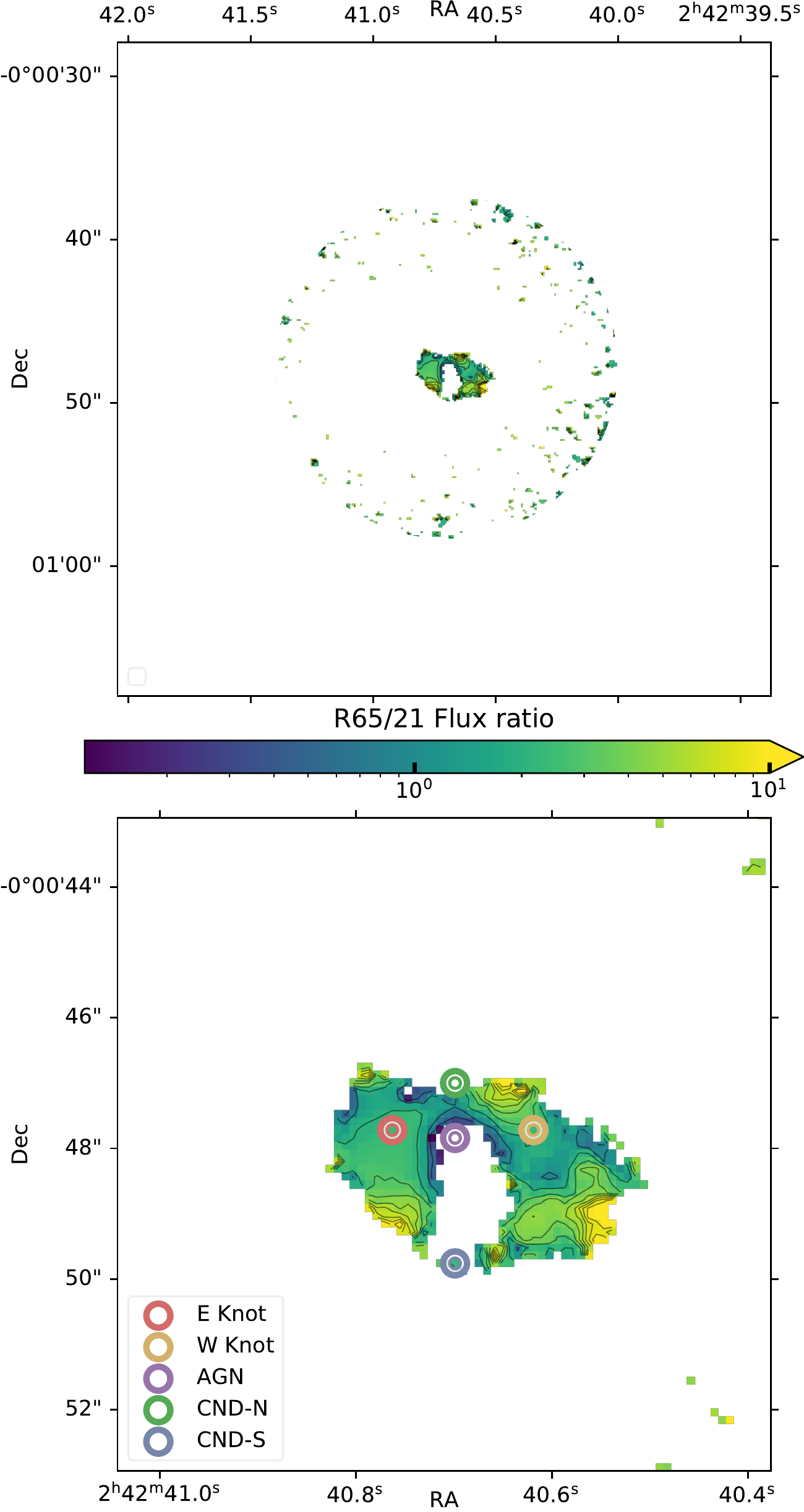}
	\caption{The line ratio CS(6-5)/CS(2-1) over the SB ring \textit{(top)} and the CND \textit{(bottom)}. The map was produced by degrading the CS(6-5) map to the same resolution as the CS(2-1) map ($0.75'' \times 0.51''$) and is truncated at the 21$''$ FOV of the CS(6-5) map. The contours correspond to 0.1, 0.5 and 1 through 10 in increments of 1.}
	\label{fig:r65/21}
\end{figure*}

\begin{figure*}
	\centering
    \includegraphics[width=12.5cm]{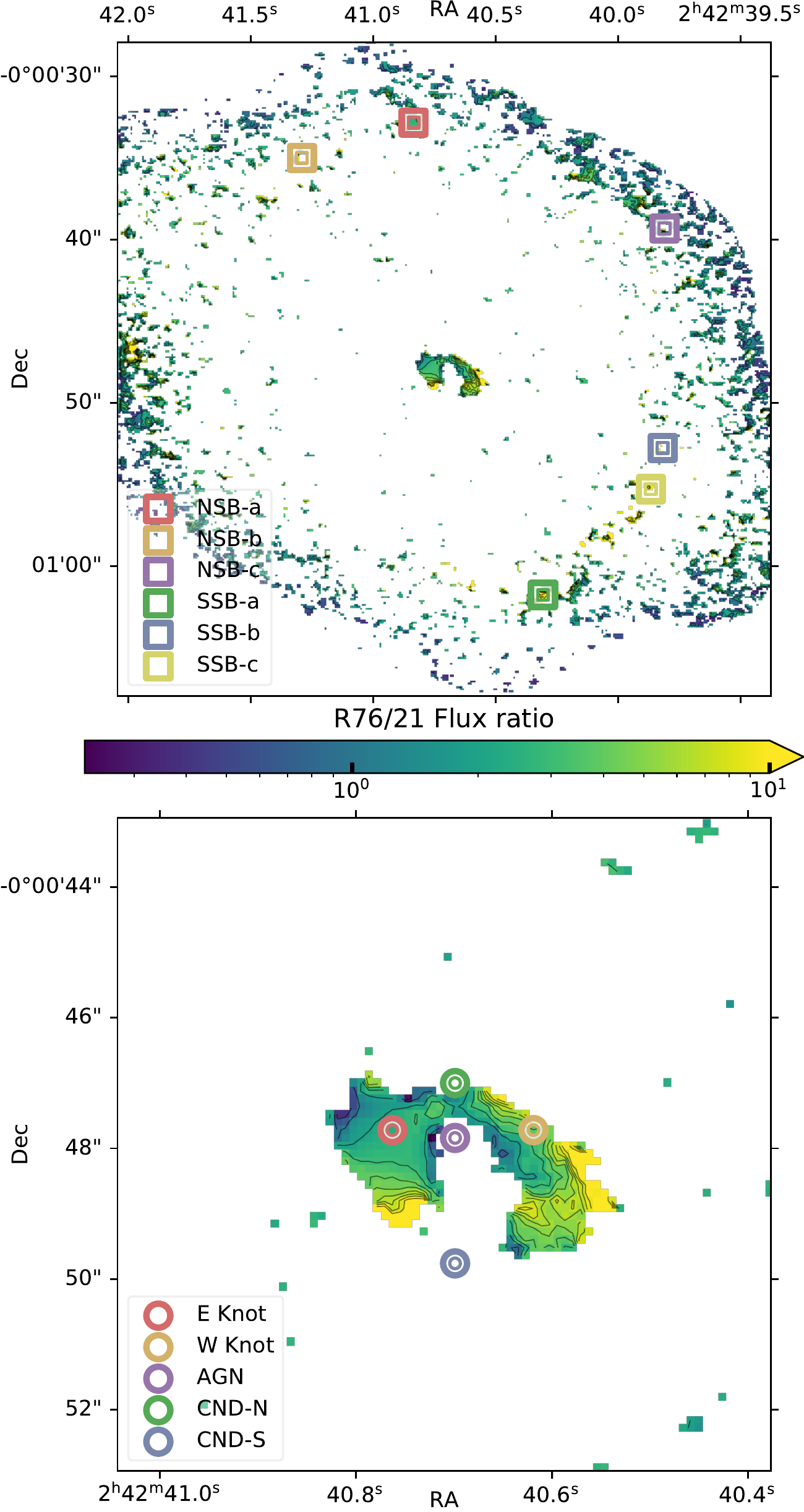}
	\caption{The line ratio CS(7-6)/CS(2-1) over the SB ring \textit{(top)} and the CND \textit{(bottom)}. The map was produced by degrading the CS(7-6) map to the same resolution as the CS(2-1) map ($0.75'' \times 0.51''$) and is truncated at the 50$''$ FOV of the CS(7-6) map. The contours correspond to 0.1, 0.5 and 1 through 10 in increments of 1.}
	\label{fig:r76/21}
\end{figure*}

Here we include the CS line ratio maps derived in this work: R32/21 (Fig. \ref{fig:r32/21}), R65/21 (Fig. \ref{fig:r65/21}) and R76/21 (Fig. \ref{fig:r76/21}). The production and analysis of these maps is detailed in Sect. 3.2 of main text. In each figure the top panel shows the map over the SB ring, truncated to the FOV of the transition covering a smaller area (in each case, this is the non CS(2-1) transition), and indicated sub-regions of interest in the SB ring. The lower panel shows the same map zoomed in to show the CND, with sub-regions within indicated.

\FloatBarrier

\section{RADEX models}
\label{sec:chi2s}

\renewcommand{\thefigure}{C\arabic{figure}}
\setcounter{figure}{0}

\begin{figure*}
	\centering
    \includegraphics[width=17cm]{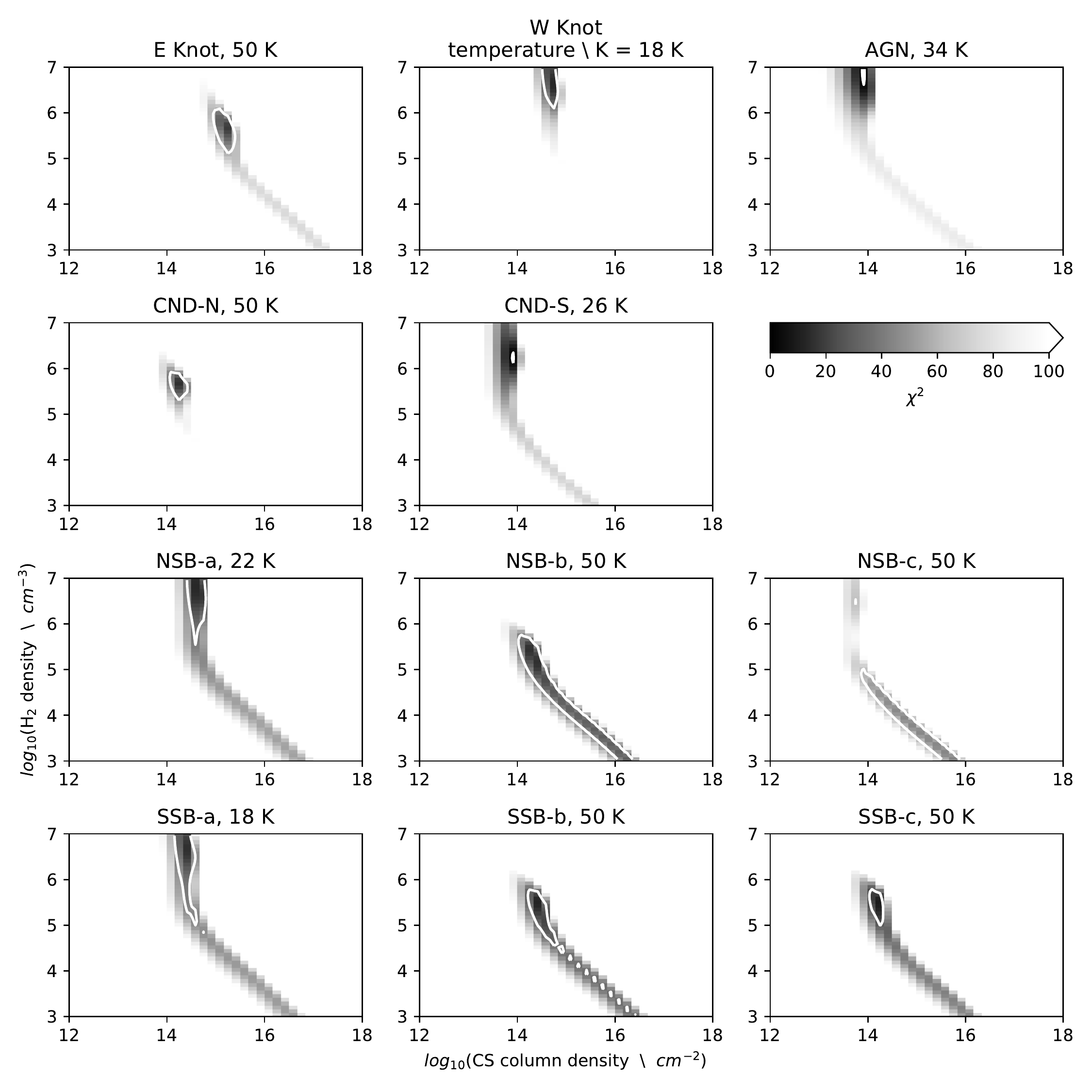}
    \caption{The $\chi^2$ fitting results for the RADEX modelling of the CS SLEDs in the various sub-regions of NGC 1068. Darker regions correspond to better fits indicated by lower $\chi^2$ values. The white contours show the limit of $\chi^2$ used to determine the best fitting models, which varies between the sub-region. The NSB-c contour corresponds to $1.5\chi_{min}^2$ due to the large value of $\chi_{min}^2$, and similarly the AGN contour to a value of $3$ due to the small value of $\chi_{min}^2$. For the remaining sub-regions the contour is placed at $3\chi^2_{min}$. Slices of constant kinetic temperature are shown, chosen to correspond to the lowest value of $\chi^2$. The exact value of temperature is given above each panel.}
	\label{fig:chi_T}
\end{figure*}

\begin{figure*}
	\centering
    \includegraphics[width=17cm]{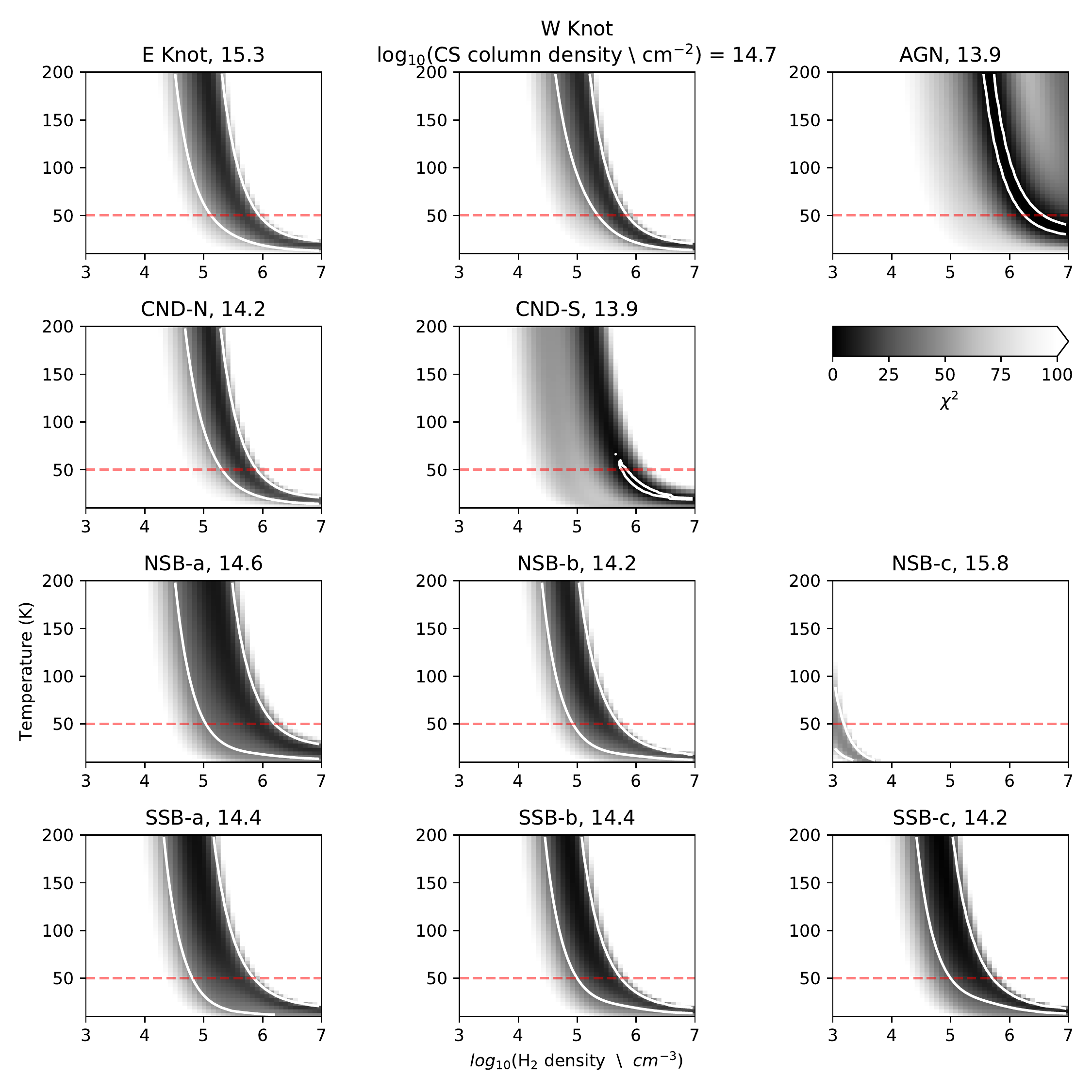}
	\caption{The $\chi^2$ fitting results for the RADEX modelling of the CS SLEDs in the various sub-regions of NGC 1068. Darker regions correspond to better fits indicated by lower $\chi^2$ values. The white contours show the limit of $\chi^2$ used to determine the best fitting models, which varies between the sub-region. The NSB-c contour corresponds to $1.5\chi_{min}^2$ due to the large value of $\chi_{min}^2$, and similarly the AGN contour to a value of $3$ due to the small value of $\chi_{min}^2$. For the remaining sub-regions the contour is placed at $3\chi^2_{min}$. Slices of constant CS column density are shown, chosen to correspond to the lowest value of $\chi^2$. The exact value of CS column density is given above each panel (as a decadic log).}
	\label{fig:chi_CS}
\end{figure*}

Here we include the $\chi^2$ distributions of the RADEX model fitting discussed in Sect. 5.1. In Fig. \ref{fig:chi_T} the slices shown have fixed temperature, and in Fig. \ref{fig:chi_CS} the CS column density is fixed. The visualisation with constant hydrogen density is included in the main text (Fig. 3). In each case, the value of the fixed variable is chosen by taking the value that gives the lowest $\chi^2$ in each sub-region.

\bsp	
\label{lastpage}
\end{document}